\documentclass[12pt]{article}

\usepackage[margin=2.5cm]{geometry}

\usepackage{amsmath,amsthm}
\usepackage{amssymb}
\usepackage{amsfonts}
\usepackage{mathtools}
\usepackage{mathrsfs}
\usepackage{times}
\usepackage{bm}
\usepackage{graphicx,natbib}
\usepackage{accents} 
\usepackage{color,hyperref}
\usepackage{enumitem}

\hypersetup{pdfauthor={Name}}

\newtheorem{theorem}{Theorem}
\newtheorem{proposition}{Proposition}
\newtheorem{lemma}{Lemma}

\theoremstyle{definition}

\renewcommand{\phi}{\varphi}
\renewcommand{\epsilon}{\varepsilon}

\newcommand{\cF}{\mathcal{F}}

\newcommand{\cL}{\mathcal{L}}

\newcommand{\cS}{\mathcal{S}}

\newcommand{\cX}{\mathcal{X}}

\newcommand{\cO}{\mathcal{O}}

\newcommand{\bR}{\mathbb{R}}

\newcommand{\pitil}{\widetilde{\pi}}
\newcommand{\xtil}{\widetilde{x}}

\newcommand{\Pkernel}{\mathsf{P}}

\newcommand{\xx}{\overline{x}}
\newcommand{\zz}{\overline{z}}
\newcommand{\uu}{\overline{u}}
\newcommand{\XX}{\overline{X}}
\newcommand{\ZZ}{\overline{Z}}
\newcommand{\UU}{\overline{U}}

\newcommand{\RR}{\overline{R}}

\newcommand{\DBPS}{{Discrete Bouncy Particle Sampler}~}

\newcommand{\bra}[1]{\left \langle #1 \right \rangle}
\newcommand{\BK}[1]{ {\left( #1 \right)} }
\newcommand{\sqBK}[1]{ {\left[ #1 \right]} }
\newcommand{\curBK}[1]{ {\left\{ #1 \right\}} }

\newcommand{\PP}{\ensuremath{\operatorname{P}}}

\newcommand{\EE}{\ensuremath{\operatorname{E}}}

\newcommand{\Var}{\ensuremath{\operatorname{Var}}}

\newcommand{\norm}[1]{\left\Vert #1 \right\Vert}
\newcommand{\tnorm}[1]{{\left\vert\kern-0.25ex\left\vert\kern-0.25ex\left\vert #1 \right\vert\kern-0.25ex\right\vert\kern-0.25ex\right\vert}}

\newcommand{\reflect}{\mathscr{R}}
\newcommand{\field}{\mathscr{F}}

\newcommand{\mylabel}[2]{#2\def\@currentlabel{#2}\label{#1}}

\newcommand{\uzeta}{\underline{\zeta}}

\newcommand{\md}{\mathsf{d}}
\newcommand{\betaE}{\mathrm{m}_2(\gamma)}

\newcommand{\fast}{\text{Fast}}
\newcommand{\suppmat}{{Supplementary Material}}

\DeclarePairedDelimiter\floor{\lfloor}{\rfloor}

\title{A Discrete Bouncy Particle Sampler}

\author{C. Sherlock$^1$ and A.H. Thiery$^2$}
\date{{\small $^1$Department of Statistics, Lancaster University, United Kingdom\\
$^2$Department of Statistics \& Applied Probability, National University of Singapore }}

\begin{document}

\maketitle

\begin{abstract}
 Most Markov chain Monte Carlo 
methods operate in discrete time and 
are reversible with respect to the target probability. Nevertheless, it is now understood 
that the use of non-reversible Markov chains can be beneficial in many
contexts. In particular, the recently-proposed Bouncy Particle Sampler 
leverages a continuous-time and non-reversible Markov process and empirically shows state-of-the-art performances when used to explore certain probability densities; however, its implementation typically requires the computation of local upper bounds on the gradient of the log target density. 

We present the Discrete Bouncy Particle Sampler, a general algorithm based upon a guided random walk, a partial refreshment of direction, and a delayed-rejection step.
We show that the Bouncy Particle Sampler can be understood as a scaling limit of a special case of our algorithm. In contrast to the Bouncy Particle Sampler, implementing the Discrete Bouncy Particle Sampler only requires point-wise evaluation of the target density and its gradient. We propose extensions of the basic algorithm for situations when the exact gradient of the target density is not available. 
In a Gaussian setting, we establish a scaling limit for the radial process
as dimension increases to infinity.
We leverage this result to obtain the theoretical efficiency of the Discrete Bouncy Particle Sampler as a function of the partial-refreshment parameter, which leads to a simple and robust tuning criterion. 
A further analysis in a more general setting 
suggests that this tuning criterion applies more generally. Theoretical and empirical efficiency curves are then compared for different targets and algorithm variations.
\end{abstract}


%
%
\section{Introduction}
\label{sec.intro}
Markov Chain Monte Carlo (MCMC) algorithms provide Monte Carlo approximations to expectations with respect to a given probability distribution, $\pi$, via an ergodic Markov chain whose invariant distribution is $\pi$.
Non-reversible Markov Chaine Monte Carlo 
samplers,
of which the Hamiltonian Monte Carlo algorithm \citep{duane1987hybrid} is
perhaps one of the most successful and widely-used examples, 
are known to enjoy desirable mixing properties in several contexts. Indeed,
several theoretical results quantify the advantages of non-reversible samplers. For example,
\cite{diaconis2000analysis} obtains rates of convergence for a 
non-reversible version of the random walk algorithm;  subsequently and
inspired by \cite{diaconis2000analysis}, \cite{chen1999lifting}
describes the best acceleration achievable through the idea of 
lifting. On a different note, 
\cite{hwang2015variance}, \cite{lelievre2013optimal}, \cite{rey2015irreversible} and \cite{duncan2016variance}
investigate and quantify the advantages offered by leveraging 
(a discretization of) a non-reversible diffusion process for computing Monte-Carlo averages: in many settings, it can be proved that the standard reversible Langevin dynamics is the worst in terms of asymptotic variances among a large class of diffusion processes that are ergodic with respect to a given target distribution. More recently, different designs of non-reversible MCMC sampler have been proposed. The Zig-Zag sampler, an instance of the large class of Piecewise-Deterministic-Markov-Processes, first obtained as a scaling limit of a lifted Metropolis--Hastings Markov chain \citep{bierkens2017piecewise}, is a continuous-time non-reversible Markov process that can be used for computing ergodic averages, and can be used for efficiently exploring Bayesian posterior distributions in the Big-Data regime \citep{bierkens2019zig}. Inspired from the physics literature \citep{peters2012rejection}, the Bouncy Particle Sampler \citep{bouchard2017bouncy} is another continuous-time non-reversible Monte-Carlo sampler that demonstrates state-of-the-art performance when used to explore certain Bayesian posterior distributions. \cite{FBPR2018} reviews the Zig-Zag sampler and the Bouncy Particle Sampler and describes some extensions. The Bouncy Particle Sampler or the Zig-Zag Sampler requires more than simple point-wise evaluations of the log-target density and its gradient: one typically needs local upper bounds on derivatives of the log-target density. Unfortunately, those bounds are unavailable or difficult to compute in many applied situations. Consequently, such continuous-time samplers cannot be directly used in these settings. This article presents a discrete-time MCMC sampler, inspired by the Bouncy Particle Sampler, that can be implemented when only point-wise evaluations of the target-density and its gradient are available. 

Our algorithm, the Discrete Bouncy Particle Sampler, is described in detail in Section \ref{sec.DBPS.algorithm}. It extends the statespace from a position to a position and a direction in the same way that the Zig-Zag and Bouncy Particle samplers do; however, our algorithm operates in discrete time and is based upon the  guided random walk of \cite{gustafson1998guided}. The guided random walk combines two reversible kernels to create a non-reversible kernel which heads in a specific direction until a rejection occurs, at which point it reverses direction. Our key addition is a particular delayed-rejection proposal \citep{TierneyMira1999}, used after any initial rejection, potentially avoiding many inefficient direction reversals. The delayed-rejection move is analogous to the bounce in the Bouncy Particle Sampler and we show that the Discrete Bouncy Particle Sampler can be viewed as a time discretization of the Bouncy Particle Sampler. An alternative discretization of the Bouncy Particle Sampler, based on
the reflective slice sampler~\citep{neal2003} is described and extended in the independent work of
\cite{VanettiBouchardDelDoucet2017}. Importantly, several interesting extensions have recently been proposed \citep{VanettiBouchardDelDoucet2017,WuRobert2017,wu2018coordinate, monmarche2019kinetic, michel2020forward} to scale and enhance this class of Piecewise-Deterministic-Markov-Processes MCMC samplers.

As with the Bouncy Particle Sampler, our algorithm can be reducible. To solve this issue, we perturb the direction vector at the end of every iteration. The size of this per-iteration perturbation has a substantial impact on the performance of the algorithm, in a similar way to the occasional, complete direction refresh of the Bouncy Particle Sampler \citep{bierkens2018high}. Analogously to the independent investigations for the Bouncy Particle Sampler in
\cite{bierkens2018high}, for the Discrete Bouncy Particle Sampler exploring a Gaussian target we use diffusion-approximation arguments to describe the limit of the radial process as dimension increases to infinity.
We then leverage this to obtain the theoretical efficiency of the \DBPS as a function of the partial-refreshment parameter, $\kappa$.  This leads to a simple and robust tuning criterion that is described in Section \ref{sec.tuningkappa}.
In more practical developments, we show that a surrogate may be substituted for the gradient of the target density when it is computationally expensive or impossible to obtain. 
Our final contribution is a construction that allows the user to choose to only calculate a fixed number of orthogonal components of the gradient vector.

Throughout the remainder of this article, the distribution of interest is referred to as $\pi(\md x)$, and is assumed to have support on the standard Euclidean space
$\cX \equiv (\bR^d, \bra{\cdot, \cdot})$ with associated norm $\norm{\cdot}$. 
This distribution is assumed to have a density with respect to Lebesgue measure, which, with a slight abuse of notation, is also denoted by $\pi$.
We set $x \wedge y = \min(x,y)$. For a distribution $\pi$ and a $\pi$-integrable function $\phi$, the quantity $\pi(\phi)$ refers to the expectation of $\phi$ under $\pi$. For $x \in \mathbb{R}$, its positive and negative part are denoted as $x_+ \geq 0$ and $x_- \geq 0$ so that $x = x_+ - x_-$. For two vectors $u,v \in \mathbb{R}^d$, their dot product is $\bra{u,v} = u_1 \, v_1 + \ldots + u_d \, v_d$. For any time $t$, $t^-$ refers to the instant in time just prior to $t$ and $t^+$ to that just after $t$.

\section{The Discrete Bouncy Particle Sampler}
\label{sec.DBPS.decription}

\subsection{Algorithm description}
\label{sec.DBPS.algorithm}
The Discrete Bouncy Particle Sampler operates on the extended state space $\cX \times \cS$, where $\cS \subseteq \cX$, and explores the extended target distribution
\begin{align*}
\pitil(\md x, \md u) = \pi(\md x)  \otimes \rho(\md u)
\end{align*}
where $\rho(\md u)$ is an auxiliary spherically symmetric distribution with support $\cS \subset \cX$. Section \ref{sec.dirn.dyn} describes several standard choices of auxiliary distributions. 
Henceforth, we will refer to the variable $x \in \cX$ as the {position} of a particle and the variable $u \in \cS$ as its {direction}.
The bounce after which the \DBPS is named enters through the operator $u \mapsto \reflect_v(u)$ that reflects the vector $u \in \cS$ with respect to the hyperplane orthogonal to the vector $v \in \cX \setminus \{0\}$,
%
%
\begin{align} 
\label{eqn.reflect}
\reflect_v(u) 
\equiv u-2 \, \frac{\bra{u,v}}{\norm{v}^2} \, v.
\end{align}
For any vector $v \in \cX \setminus \{0\}$, the reflection operator $\reflect_v$ is an involution $\reflect_v \circ \reflect_v(u)  = u$ that preserves norms. This remark underlies the proof of correctness of the Discrete Bouncy particle Sampler whose details are presented in the \suppmat.
As discussed in \cite{michel2020forward}, it is possible to rely on more general reflection operators. Most of the methods developed in this text extends to these variants, although we concentrate on Equation \eqref{eqn.reflect} for ease of exposition. The Discrete Bouncy Particle Sampler relies on a non-vanishing vector field
\begin{align*}
\field: \cX \to \cX \setminus \{0\}.
\end{align*}
In practice, this vector field is either chosen as $\field(x) = \nabla \log \pi(x)$, replaced by an arbitrary modification when the gradient vanishes, or as an approximation of it, as described in Section \ref{sec.approxgrad}. 
The quantity $\reflect_{\field(x)}(u)$ represents the resulting direction when a particle with incoming direction $u$ performs an elastic bounce off the hyperplane orthogonal to the vector $\field(x)$. 

The Discrete Bouncy Particle Sampler deterministically cycles through two Markovian transitions that leave the extended target distribution $\pitil$ invariant, {(1)} a {Position Update}, with a possible {Direction Reflection} {(2)} a {Direction Refreshment}. The resulting scheme is, in general, non-reversible. For the direction refreshment, one can choose any Markov transition kernel $\Pkernel_\rho(u, \md u')$ that leaves the auxiliary distribution $\rho$ invariant. For a discretization parameter $\delta > 0$, and a current state $(x_k, u_k) \in \cX \times \cS$, the algorithm proceeds as follows.
\begin{enumerate}
%
%
\item \label{step.DR}
\textsc{Position Update}:
Generate a proposal $(x',u')=(x_k+ \delta \, u_k, u_k)$. With {position update} probability
\begin{align} \label{eq.alpha.first.step}
\alpha_{\mathrm{pu}}(x_k, u_k)\equiv 1 \, \wedge \, \frac{ \pi(x') }{ \pi(x_k) },
\end{align}
set $(\hat{x}_{k},\hat{u}_{k})=(x',u')$ and go to Step \ref{step.direction.randomization}. Otherwise, proceed to Step \ref{step.reflection}.
\item \label{step.reflection}
\textsc{Direction Reflection}:
consider $u''= \reflect_{\field(x')}(u')$ and $x'' = x' + \delta \,u''$. With {direction reflection} probability
%
%
\begin{align} 
\label{eq.alpha.DR}
\alpha_{\mathrm{dr}}(x_k, u_k)\equiv 1 \, \wedge \, 
\curBK{
\frac{ 1-\alpha_{\mathrm{pu}}(x'',-u'') }{1-\alpha_{\mathrm{pu}}(x_k,u_k)} \times \frac{\pi(x'')}{\pi(x_k)} },
\end{align}
set $(\hat{x}_{k},\hat{u}_{k})=(x'',u'')$. Otherwise, negate the direction by setting $(\hat{x}_{k},\hat{u}_{k})=(x_k,-u_k)$.
%
%
%
\item \label{step.direction.randomization}
\textsc{Direction Refreshment}: Set $(x_{k+1}, u_{k+1}) = (\hat{x}_k, U)$ where $\PP(U \in A) = \Pkernel_\rho(\hat{u}_k, A)$.
\end{enumerate}

By construction, the Direction Refreshment step preserves the extended target distribution. The fact that the combination of the Position Update and  Direction Reflection steps also preserves the extended distribution is discussed in the \suppmat. This can be understood as a slight generalization of the standard delayed rejection mechanism \citep{TierneyMira1999} when applied to a deterministic and volume preserving proposal. A similar scheme was proposed independently in \cite{VanettiBouchardDelDoucet2017}. Furthermore, and importantly for Section \ref{sec.approxgrad}, the algorithm remains valid if the deterministic vector field is replaced by a randomized version of it. The proof of correctness is identical to the deterministic case and is briefly discussed in the \suppmat.

A simple thought experiment, such as considering a target density  $\pi$ with spherically-symmetric contours and $\Pkernel_\rho(u, \md u')= \delta_{u}(\md u')$, shows that, as with the Bouncy Particle Sampler, the Discrete Bouncy Particle Sampler can be reducible. The choice and tuning of the direction refreshment operator $\Pkernel_\rho$ is consequently important in practice and is discussed at length in the sequel.

\subsection{Direction dynamics}
\label{sec.dirn.dyn}
The choice of the auxiliary distribution $\rho(\md u)$ and the update operator $\Pkernel_\rho$, which we now allow to depend on a tuning refreshment parameter, $\kappa$, has a major influence on the efficiency of the resulting Discrete Bouncy Particle Sampler. 
Two standard choices for the isotropic distribution $\rho$ are the uniform distribution $\rho_S$ on the unit sphere of $\mathbb{R}^d$ and the centred Gaussian distribution $\rho_G$ with covariance $(1/d) \, I_d$. 
The scaling of the covariance matrix implies that $\int \|u\|^2 \, \rho_{G}(\md u) = \int \|u\|^2 \, \rho_{S}(\md u)  = 1$, which ensures that the effect on the algorithm of the tuning parameters $\delta$ and $\kappa$ are comparable across the different auxiliary distributions. As will be described in Section \ref{sec.continuous.limit}, it is convenient to think of $\Pkernel_\rho(u, \md u')$ as the discretization between time $t$ and $t + \delta$ of a continuous-time Markov process $\{V_t\}_{t \geq 0}$ that leaves $\rho(\md u)$ invariant, $\Pkernel_\rho(u, \md u') = \PP(V_{t + \delta} \in \md u' | V_t = u)$. We now describe several standard ways of generating a random variable $U$ that, conditioned on a current direction $u \in \cS$, is approximately distributed as $\Pkernel_\rho(u, \md u')$.

\begin{itemize}
\item \textsc{Full Refresh}: for $\rho=\rho_S$ or $\rho=\rho_G$ and an update rate $\kappa > 0$ the Markov process with generator $\mathcal{L}^{(V)} \phi(u) = \kappa \, \BK{\rho(\phi) - \phi(u)}$ 
completely refreshes the direction at rate $\kappa$ and has a mixing time of $\mathcal{O}(1/\kappa)$. For a time discretization parameter $0< \delta < 1$, set
\begin{align*}
U = B_{\kappa}^{\delta} \, u + (1-B_{\kappa}^{\delta}) \, \xi
\end{align*}
where $\xi \sim \rho$ and $B_{\kappa }^{ \delta}$ is a Bernoulli random variable with $\PP(B_{\kappa}^{\delta}=1)=\exp(-\kappa \, \delta)$.

\item \textsc{Ornstein-Uhlenbeck refresh}: for $\rho=\rho_G$ and an update rate $\kappa > 0$, the Ornstein-Uhlenbeck process $\md V_t = -(\kappa/2) \, V_t \,\md t + (\kappa / d)^{1/2} \, \md W$ leaves $\rho_G$ invariant and has a mixing time of $\mathcal{O}(1/\kappa)$. Set
\begin{align} \label{eq.OU.update}
U = \alpha \, u + (1-\alpha^2)^{1/2} \, \xi
\end{align}
for $\xi \sim \rho_G$ and $\alpha = \exp(-\kappa \, \delta / 2)$.

\item \textsc{Brownian motion on the unit sphere}: for $\rho=\rho_S$ and an update rate $\kappa > 0$, consider the Brownian motion on the unit sphere in $\mathbb{R}^d$. It is described by the stochastic differential equation
\begin{align} \label{eq.brownian.SDE}
\md V_t = - \frac{\kappa}{2} \, V_t \, \md t + \{\kappa / (d-1)\}^{1/2} \, P_{\perp}(V_t) \, \md W
\end{align}
where $W$ is a standard Brownian motion in $\mathbb{R}^d$ and $P_{\perp}(V_t) \in \mathbb{R}^{d,d}$ is the orthogonal projection on the hyperplane orthogonal to $V_t$. The Brownian motion on the unit sphere leaves $\rho_S$ invariant and has a mixing time of $\mathcal{O}(1/\kappa)$. One can define $U$ by normalizing an Ornstein-Uhlenbeck update \eqref{eq.OU.update},
\begin{align*}
U = \frac{ \alpha \, u + (1-\alpha^2)^{1/2} \, \xi }{\|\alpha \, u + (1-\alpha^2)^{1/2} \, \xi\|},
\end{align*}
for $\xi \sim \rho_G$ and $\alpha = \exp(-\kappa \, \delta / 2)$. 
\end{itemize}

\subsection{Continuous-time limit}
\label{sec.continuous.limit}
In this section we show that, as $\delta \to 0$, the Discrete Bouncy Particle Sampler converges to a well defined continuous-time and piecewise-continuous Markov process. This result clarifies the role of the reflection operator $\reflect_{\field(x)}$ and explains the connection between the Discrete Bouncy Particle Sampler and the continuous time Bouncy Particle Sampler. 
For any time discretization parameter $\delta > 0$, consider a Discrete Bouncy Particle Sampler chain $\{(x^{\delta}_k, u^{\delta}_k)\}_{k \geq 0}$ with update operator $\Pkernel^\delta_\rho(u, \md u')$. The superscript $\delta$ indicates, as will be made clearer in Assumption ${A3}$ stated below, that the update operator $\Pkernel^\delta_\rho(u, \md u')$ is obtained as the discretization of a Markov process $\{V_t\}_{t \geq 0}$ between time $t$ and $t+\delta$. For a time horizon $T > 0$, consider the continuous time process $\zz^\delta_t = (\xx^\delta_t, \uu^\delta_t)$ defined on the interval $[0,T]$ by setting
\begin{align*}
\zz^\delta_{k \delta} = (\xx^\delta_{k \delta}, \uu^\delta_{k \delta}) = (x^{\delta}_k, u^{\delta}_k) \in \cX \times \cS
\end{align*}
for any integer $0 \leq k \leq T/\delta$  and linearly interpolation in between. 
The main result of this section, Proposition \ref{prop.delta.zero.limit}, whose proof is presented in the \suppmat, relies on the following regularity assumptions.
%
%
\begin{description}
\item[A1] The function $x \mapsto \log \pi(x)$ is twice differentiable with a bounded second derivative.
\item[A2] The vector field $\field: \cX \to \cX \setminus \{0\}$ is continuous. 
\item[A3] There exists a continuous time Markov process $\{ V_t \}_{t \geq 0}$ with generator $\mathcal{L}^{(V)}$ such that, for any time discretization parameter $\delta > 0$, the transition kernel $\Pkernel^{\delta}_\rho$ describes the transition of the Markov process $V$ in the sense that $\Pkernel^{\delta}_\rho(u, \md u') = \PP(V_{t+\delta} \in \md u' | V_t = u)$. We assume that the trajectories of the Markov process $\{ V_t \}_{t \geq 0}$ are almost surely continuous.
\end{description}
\begin{proposition} \label{prop.delta.zero.limit}
Let Assumptions {A(1-2-3)} hold and consider a fixed time horizon $T>0$. As $\delta \to 0$, the sequence of continuous time processes $\zz^\delta_t = (\xx^\delta_t, \uu^\delta_t)$ converges weakly in the Skorokhod topology to the bivariate Markov process $\ZZ_t = (\XX_t, \UU_t)$ with generator
\begin{align} \label{eq.generator.Z}
\begin{aligned}
\mathcal{L}^{(\ZZ)} & \, \phi(x,u) = 
 \mathcal{L}^{(V)}\phi(x,u) + \bra{u , \nabla_x \, \phi(x,u)} \\
 & + \lambda(x,u) \, \big( [ \mathcal{A}(x,u) \, \phi\{x,\reflect_{\field(x)}(u)\} + \{1-\mathcal{A}(x,u)\} \, \phi(x,-u) ]- \phi(x,u) \big)
\end{aligned}
\end{align}
with rate $\lambda(x,u) \equiv \bra{-\nabla \log \pi(x), u}_+$ and acceptance probability %
%
\begin{align} \label{eq.limiting.acceptance}
\mathcal{A}(x,u) \equiv 1 \wedge \frac{\lambda\{x, -\reflect_{\field(x)}(u)\}}{\lambda(x,u)} \in [0,1].
\end{align}
\end{proposition}
The limiting Markov process $\ZZ_t = (\XX_t, \UU_t)$ with generator \eqref{eq.generator.Z} evolves according to the dynamics 
\begin{align} \label{eq.X.dynamics}
\md \XX_t = \UU_t \, \md t,~~~\md \UU_t = \md V_t,
\end{align}
in between events that arrive at rate $\lambda(\XX_t, \UU_t)$. When such an event is triggered, the direction is reflected, i.e. $\UU_t = \reflect_{\field(X_{t^-})}(\UU_{t^-})$, with probability $\mathcal{A}(\XX_{t^-}, \UU_{t^-})$, and completely reversed, i.e. $\UU_t = -\UU_{t^-}$, with probability $1-\mathcal{A}(\XX_{t^-}, \UU_{t^-})$. Possible choices of Markovian dynamics with generator $\mathcal{L}^{(V)}$ in Assumption {A2}  are detailed in Section \ref{sec.dirn.dyn}. In the case when $\field(x) = \nabla \log \pi(x)$ and $\cL^{(V)}\phi(u) = \kappa \, \curBK{\rho(\phi) - \phi(u)}$, for a fixed refreshment rate $\kappa > 0$, the limiting Markov process  is the standard Bouncy Particle Sampler \citep{bouchard2017bouncy}. The interested reader is referred to \cite{VanettiBouchardDelDoucet2017,WuRobert2017} for other interesting generalizations

In order to understand the influence of the vector field $\field$, it is instructive to study the limiting acceptance probability \eqref{eq.limiting.acceptance}. The limiting process $\ZZ$ is rejection free, i.e. never backtracks, if for any $(x,u) \in \cX \times \cS$ we have that $\lambda\{x, -\reflect_{\field(x)}(u)\} = \lambda(x,u)$. It is readily seen that this condition is equivalent to choosing $\field(x)$ proportional to $\nabla \log \pi(x)$ for any $x \in \cX$ where this quantity does not vanish. In other words, any other choice of vector field $\field$ leads to a limiting process that is not rejection-free. Section \ref{sec.approxgrad} describes ways to efficiently approximate this optimal choice when evaluating $\nabla \log \pi(x)$ is not computationally efficient.

\subsection{Approximate reflections}
\label{sec.approxgrad}
As described in the previous section, vector fields that lead to a rejection-free algorithm in the limit $\delta \to 0$ are such that $\field(x)$ is proportional to $\nabla \log \pi(x)$ for all $x \in \cX$ where this quantity is non-zero. When computing the gradient of the log-density is not computationally feasible, one can instead use a vector field $\field$ that only approximates $\nabla \log \pi$, necessarily paying the price of a non-zero probability for the limiting algorithm to backtrack. Another strategy, similar to the one presented in \cite{fielding2011efficient}, consists in choosing $\field(x)$ as the gradient of an approximate surrogate target distribution.

Assume that the Discrete Bouncy Particle Sampler stands at $(x,u) \in \cX \times \cS$ and that a {Direction Reflection} is attempted. 
The exact gradient $g(x) \equiv \nabla \log \pi(x)$ being unavailable, one can instead numerically evaluate $g(x) \in \mathbb{R}^d$ along a set of $n_{\text{cpt}} \leq d$ random directions described by a set of $n_{\text{cpt}}$ mutually orthogonal unit vectors $\uzeta \equiv (\zeta_1,\dots,\zeta_{n_{\text{cpt}}})$ generated from a distribution $G_x(\md \uzeta)$ that may depend on the current position $x \in \cX$ but not the current direction $u \in \cS$. As described in the \suppmat, the fact that the distribution $G_x(\md \uzeta)$ does not depend on the current direction ensures that the resulting algorithm has the correct invariant distribution. A standard choice consists in orthonormalizing a set of $n_{\text{cpt}}$ vectors generated from an isotropic Gaussian distribution. The approximate gradient is then defined as $\widetilde{g}(x) = \sum_{i}^{n_{\text{cpt}}} \bra{g(x), \zeta_i} \, \zeta_i$, where each coefficient$\bra{g(x), \zeta_i}$ can be evaluated numerically \citep{ramm2001stable}. In other words, $\widetilde{g}(x)$ is the orthogonal projection of the vector $g(x)$ onto the plane $V(\uzeta) \equiv \mathrm{span}(\zeta_1,\dots,\zeta_{n_{\text{cpt}}})$. Decomposing the direction as $u = u^{\perp} + u^{\parallel}$ with 
$u^{\parallel} \in V(\uzeta)$ and $u^{\perp} \in V(\uzeta)^{\perp}$, the following two modified bounce operators $u \mapsto u'$ both lead to a correct algorithm:
\begin{enumerate}
\item The updated direction $u' \in \cS$ can be defined as the reflection with respect to the hyperplane orthogonal to $\widetilde{g}(x)$, i.e. $u' = \reflect_{\tilde{g}(x)}(u)$. This update can also be expressed as
\begin{align*}
u' &= \reflect_{\tilde{g}(x)}(u)=u^{\perp} + \reflect_{\tilde{g}(x)}(u^{\parallel}).
\end{align*}
\item One can also completely reflect the component of $u \in \cS$ that is orthogonal to the plane $V(\uzeta)$. In other words, the updated direction $u'$ is defined as
\begin{align} \label{eq.reflection.tweak}
u' = -u^{\perp} + \reflect_{\tilde{g}(x)}(u^{\parallel}).
\end{align}
\end{enumerate}
While both reflection operators are valid, we have empirically found that the operator \eqref{eq.reflection.tweak} leads to better mixing properties.
\subsection{Preconditioning}
\label{sec.precond}
A general target may have very different length scales in different directions and, just as with Metropolis-Hastings algorithms such as the random walk Metropolis or the Metropolis-Adjusted Langevin algorithm \citep{RobRos2001}, the efficiency can be improved, often by several orders of magnitude, by preconditioning. Consider an invertible matrix $\Gamma \in \bR^{d,d}$ and define the whitened variable $\xtil$ implicitly defined as $x = \Gamma \, \xtil$. Instead of using the Discrete Bouncy Particle Sampler for exploring a target density $\pi(x)$, one can instead explore the whitened density $\tilde{\pi}(\xtil) \propto \pi(\Gamma \, \xtil)$. 
See \cite{Pakmanetal2017} for an analogous description of preconditioning for the BPS.
Typically, the transformation $\xtil \mapsto  \Gamma \xtil$ is chosen such that the transformed density $\tilde{\pi}$ is as isotropic as possible. A standard strategy is to choose $\Gamma$ such that $\Gamma \Gamma^\top$ is an approximation of the covariance matrix of the target distribution $\pi$, or of the negative inverse Hessian of $\log \pi$ at a mode. 
%
%
%
\section{Algorithm tuning through diffusion approximation}
\label{mainsec.diffusion}

\subsection{Diffusion Limit} 
\label{seq.diffusion}
In this subsection, we derive a diffusion approximation, in a Gaussian setting, for the log-target process in the high-dimensional regime $d \to \infty$, as the discretization parameter $\delta \to 0$. We are interested in understanding the mixing properties of the process $k \mapsto \log \pi^d(x^{d,\delta}_k)$ when $\{(x^{d,\delta}_k, u^{d,\delta}_k)\}_{k \geq 0}$ is a Discrete Bouncy Particle Sampler chain targeting the $d$-dimensional Gaussian distribution $\pi^d$. One could also study the mixing properties of any other functional of the Markov chain: for a fixed dimension $D \geq 1$ and a sequence of functions $F_d: \mathbb{R}^d \to \mathbb{R}^D$ indexed by $d \geq 1$, one can investigate the properties of the stochastic process $k \mapsto F_d(x^{d, \delta}_k)$. For example, \citet{deligiannidis2018randomized} proves scaling limits of a finite and fixed set of coordinates, which corresponds to the projection operator $F_d(x_1, \ldots, x_d) = (x_1, \ldots, x_D)$. Our choice  $F^d(x) = \log \pi^d(x)$ is motivated by the empirical observation that in many scenarios, when the  Discrete Bouncy Particle Sampler is employed, the mixing of the log-target process is much slower than the mixing of any given coordinate. Similar empirical observations are discussed in  \cite{terenin2018piecewise}, and \cite{bierkens2018high} proves different diffusion limits for the standard Bouncy Particle Sampler when full refreshments are used to ensure ergodicity. Using entirely different techniques, \cite{andrieu2018hypercoercivity} obtains insights into the scaling of the Bouncy Particle Sampler and more general Piecewise Deterministic Markov Processes. The expression we obtain for the limiting process allows us in Section \ref{sec.tuningkappa} to formulate robust strategies for tuning the refreshment parameter $\kappa > 0$.
Consider a Discrete Bouncy Particle Sampler with  reflection $\reflect_{\field(x)}$ where $\field(x) = \nabla \log \pi^d$ and when the refreshment updates are obtained as discretization of a Brownian motion on the unit sphere, as described in Section \ref{sec.dirn.dyn}. 
We concentrate on the case where the target distribution $\pi_d$ is a centred $d$-dimensional Gaussian density with covariance $\sigma^2 \, \mathrm{I}_d$, for a fixed standard deviation $\sigma > 0$. Thanks to the rotational symmetry of $\pi_d$, all the bounce attempts are accepted and studying the log-target process is equivalent to studying the radial process $k \mapsto \|x^{d,\delta}_k\|$. Proposition \ref{prop.delta.zero.limit} identifies, for a fixed dimension $d \geq 1$, the scaling limit as $\delta \to 0$ of the Discrete Bouncy Particle Sampler chain $\{x^{d,\delta}_k, u^{d, \delta}_k \}_{k \geq 0}$. For concreteness, we denote this limiting continuous-time Markov process, whose generator is described in Proposition \ref{prop.delta.zero.limit}, as $(\XX^d_t, \UU^d_t)$. 
Now, for investigating the high-dimensional regime $d \to \infty$, note that for a sequence of random variables $X^d \sim \pi_d$, the sequence $\|X^d\| - \sigma \, d^{1/2}$ converges in distribution towards a centred Gaussian distribution with variance $\sigma^2/2$. We consequently define the shifted process
%
%
\begin{align} \label{eq.radial.process}
R^d_t = \| \XX^d_{d \times t} \| - \sigma \, d^{1/2}.
\end{align}
Note that, in the definition of the process $R_t^d$, time has been accelerated by a factor of $d$. Proposition \ref{prop.diffusion.limit} stated below shows that, in order to observe a non-degenerate scaling limit as $d \to \infty$, this acceleration factor is the correct one. As will be demonstrated, the mixing properties of $R^d_t$ are closely related to the mixing properties of the scalar jump-diffusion $\{ \theta^\kappa_t \}_{t \geq 0}$ with generator
\begin{align} \label{eq.generator.fast}
\cL^{(\kappa, \sigma)} = \frac{\kappa}{2} \, \cL^{K} + \frac{1}{\sigma} \cL^{J}
\qquad \textrm{where} \qquad
\left\{
\begin{aligned}
\cL^{(K)}\phi(\theta) &= -\theta \, \phi'(\theta) + \phi''(\theta)\\
\cL^{(J)}\phi(\theta) &= \phi'(\theta) + \theta_+ \, \{\phi(-\theta) - \phi(\theta)\}.
\end{aligned}
\right.
\end{align}
The operator $\cL^{(K)}$ is the generator of an Ornstein-Uhlenbeck process that is reversible with the standard Gaussian density. Similarly, $\cL^{(J)}$ is the generator of the Markov process with unit drift and reflections $\theta \mapsto -\theta$ that occur at rate $\theta_+ = \max(0,\theta)$. It can readily be checked that this process also leaves the standard Gaussian distribution invariant. Combining these two facts show that the process $\theta^\kappa_t$ also leaves the standard Gaussian distribution invariant. We denote by $V_\sigma(\kappa)$ the asymptotic variance of ergodic averages along $\theta^\kappa_t$ defined as
%
%
\begin{align} \label{eq.velocity.function}
V_\sigma(\kappa) = \lim_{T \to \infty} \; \Var
\BK{ \frac{1}{\sqrt{2 \, T}} \int_0^T  \theta^{\kappa}_t \, \md t }.
\end{align}
There is no closed form expression for the quantity $V_\sigma(\kappa)$ but it can easily be approximated numerically, as displayed in Figure \ref{fig.theory}. Note that $V_\sigma(\kappa) \to 0$ as $\kappa \to 0$ and $\kappa \to \infty$. Proposition \ref{prop.diffusion.limit}, whose proof can be found in the \suppmat, shows that the asymptotic variance $V_\sigma(\kappa)$ dictates the mixing rate of the log-target process. The higher the asymptotic variance $V_\sigma(\kappa)$, the faster the mixing of the radial process.
%
%
\begin{proposition} \label{prop.diffusion.limit}
Let $T>0$ be a finite time horizon. For a fixed refreshment parameter $\kappa > 0$, the sequence of accelerated processes $R^d$ defined in Equation \eqref{eq.radial.process} converges weakly as $d \to \infty$ in $C([0,T], \mathbb{R})$ to the Ornstein-Uhlenbeck process
\begin{align} \label{eq.limiting.diffusion}
d \RR_t = - \frac{V_\sigma(\kappa) }{\sigma^2 / 2}\, \RR_t \, dt + \{2 \, V_\sigma(\kappa)\}^{1/2} \, dW.
\end{align}
The velocity function $V_\sigma(\kappa)$ is defined in Equation \eqref{eq.velocity.function}.
\end{proposition}
The process \eqref{eq.limiting.diffusion} is an Ornstein-Uhlenbeck that is reversible with respect to the centred Gaussian distribution with variance $\sigma^2/2$.
Since the Ornstein-Uhlenbeck \eqref{eq.limiting.diffusion} has a mixing time of order $\mathcal{O}(1)$, this indicates that in the high-dimensional regime $d \to \infty$ and $\delta \to 0$, one can expect \citep{roberts2016complexity} the log-target process to mix on a time scale of order $\mathcal{O}(d / \delta)$. When implementing the Discrete Bouncy Particle Sampler in practice, the parameter $\delta$ should be chosen small enough to guarantee that the acceptance rate remains high-enough, but not smaller. Similar guidelines for the Hamiltonian Monte-Carlo method are described in \cite{betancourt2014optimizing}. 
The optimal tuning of the parameter $\delta$, and study of its dependence with respect to the dimensionality of the target distribution, is beyond the scope of this article. Instead, we concentrate on the tuning of the refreshment parameter $\kappa$. When optimising the mixing of the log-target process, we observe empirically that the tuning of the parameter $\kappa$ is insensitive to the value of $\delta$. This is in part because whatever the value of $\delta$, as long as it is sufficiently small, the log-target process is an approximation of the limiting diffusion \eqref{eq.limiting.diffusion} (see also Section \ref{sec.tuningkappa}); empirical evidence that this insensitivity continues to hold for large $\delta$ is provided in Section \ref{sec.sim.kappa.insens.to.delta}.

%
%
\subsection{Tuning of the refreshment parameter $\kappa$}
\label{sec.tuningkappa}
\begin{figure}[t]
\begin{center}
    \includegraphics[scale=0.5]{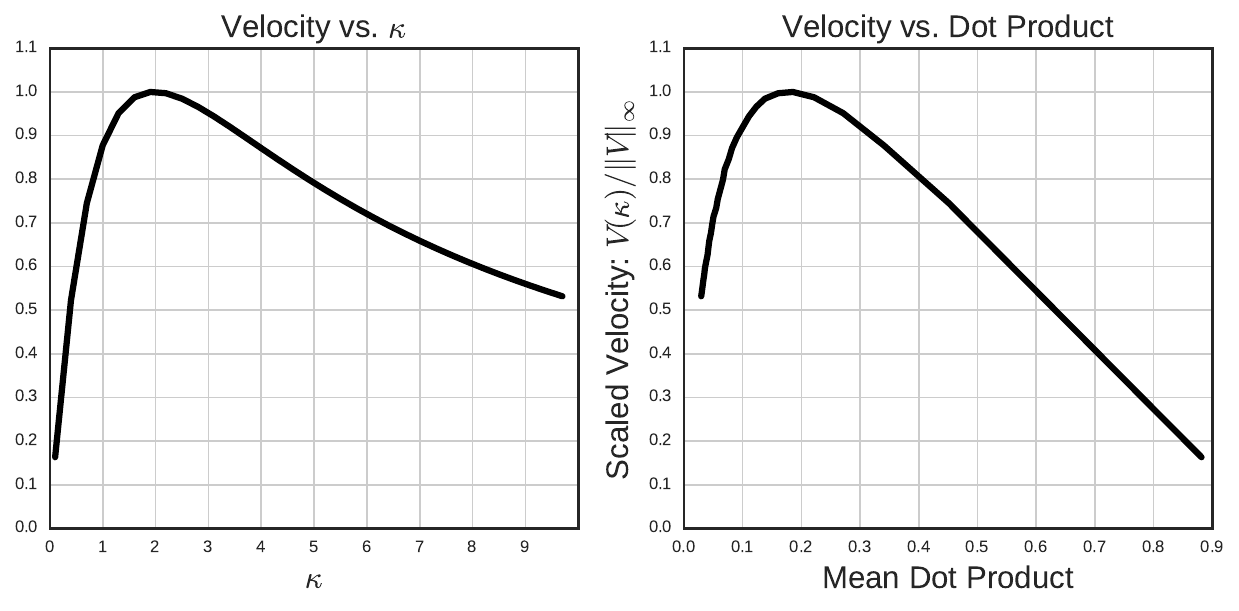}
    \end{center}
\caption{Velocity $V(\kappa)$ for $\sigma=1$ as a function of $\kappa>0$ (Left) and as a function of the mean dot product (Right). The scale of the velocity function being irrelevant, only the rescaled velocity function $\kappa \mapsto V(\kappa) / \|V\|_\infty$ is depicted.}
\label{fig.theory}
\end{figure}
\label{sec.kappa.tuning}
The limiting diffusion \eqref{eq.limiting.diffusion} obtained in Section \ref{seq.diffusion} indicates that, for an isotropic Gaussian target distribution with marginal variance $\sigma^2$ and in the regime $d \to \infty$, optimising the efficiency of the Discrete Bouncy Particle Sampler is achieved by choosing a refreshment parameter $\kappa>0$ that maximizes the velocity $V_\sigma(\kappa)$. In other words, for a given marginal variance $\sigma^2>0$, the optimal refreshment rate $\kappa_\star(\sigma)$ is given by
\begin{align*}
\kappa_\star(\sigma) = \mathsf{argmax} \; \curBK{ \kappa \; \mapsto \; V_\sigma(\kappa) }.
\end{align*}
Furthermore, a change of time argument immediately shows that $\kappa_\star(\sigma) = \kappa_\star / \sigma$  
with $\kappa_\star \equiv \kappa_\star(\sigma=1)$. 
In practice, the variance parameter $\sigma^2>0$ is not known so that the optimal refreshment parameter is not directly accessible.
%
To make progress, denote by $\{ \tau_j \}_{j \geq 1}$ the (strictly increasing) sequence of time indices at which {Direction Reflection} events are attempted (and always accepted in the Gaussian setting). We denote by $u_{\tau^-_j} \in \cS$ the direction right before a reflection event, and by $u_{\tau^+_j} \in \cS$ the direction right after the reflection. For tuning purposes, we propose to monitor the dot product $\beta \in [-1,1]$ between the direction vectors right after and before the {Direction Reflection} attempts,
\begin{align}\label{eqn.betaisdot}
\beta = \bra{u_{\tau^+_j}, u_{\tau^-_{j+1}}}.
\end{align}
For a Discrete Bouncy Particle Sampler evolving at stationarity, as $\delta \to 0$ and for any fixed dimension $d \geq 2$, consider the distribution $\mu^{d}(\md \beta; \kappa,\sigma)$ of these dot products.
One can readily check that if $\{ x^{d,\delta}_k, u^{d, \delta}_k \}_{k \geq 0}$ is Discrete Bouncy Particle Sampler chain with parameters $\kappa,\delta>0$ exploring the centred $d$-dimensional Gaussian with marginal standard deviation $\sigma$ then, for any scaling factor $s > 0$, the Markov chain defined as $\{  s \times x^{d,\delta}_k, u^{d, \delta}_k \}_{k \geq 0}$ is also Discrete Bouncy Particle Sampler chain, with refreshment parameter $\kappa / s$ and time discretization parameter $ s \, \delta>0$, exploring the centred $d$-dimensional Gaussian with marginal standard deviation $s \, \sigma$.
It follows that $\mu^d(\md \beta,\kappa,\sigma) = \mu^d(\md \beta; \kappa / s, s \, \sigma)$ for any scaling factor $s > 0$. Consequently, since $\kappa_\star(\sigma) = \kappa_\star / \sigma$, the distribution $\mu^d(\md \beta; \kappa_\star(\sigma),\sigma) \equiv \mu_{\star}^{d}(\md \beta)$ does not depend on the standard deviation $\sigma$. 
It is straightforward to numerically estimate the average dot product at optimality, 
\begin{align*}
\beta_\star \equiv \lim_{d \to \infty} \; 
\int_{-1}^1 \beta \, 
\mu_\star^d(\md \beta)
\approx 0.2.
\end{align*}
Figure \ref{fig.theory} illustrates this optimality result.
Very low values $\beta \ll \beta_\star$ indicate that the directions are updated too frequently, leading to an inefficient random-walk behaviour. High values $\beta \approx 1$ indicate that the directions are not updated frequently enough, leading to an inefficient exploration of the state space. The case $\beta = 1$ corresponds to the case when the direction are not updated at all, which is known in the Gaussian case to lead to a reducible Markov Chain with an incorrect invariant distribution.
For tuning the refreshment parameter $\kappa>0$ of a general Discrete Bouncy Particle Sampler, we consequently propose to estimate empirically the expectation at stationarity of the quantity $\beta$ in \eqref{eqn.betaisdot}. Let $\{ \tau_j \}_{j \geq 1}$ now denote the realization of the sequence of indices at which direction reflection attempts occur. A direction reflection attempt is accepted with probability \eqref{eq.alpha.DR}, otherwise the direction is negated. 
We define
\begin{align*}
\widehat{\beta}=\frac{1}{J} \sum_{j=1}^{J} \bra{u_{\tau_j^+}, u_{\tau_{j+1}^-}}
\end{align*}
and choose $\kappa > 0$ so that this quantity approximately equals its optimal value $\beta_\star \approx 0.2$. Just as with the estimation of acceptance rates when tuning the scaling of various algorithms \citep{RobRos2001}, and unlike the Effective Sample Size itself that is notoriously difficult to reliably estimate, the mean dot product can be estimated accurately from short MCMC runs. Importantly, and as described in Section \ref{sec.simulations}, we have found this tuning procedure to be robust with respect to departure from Gaussianity and to approximately hold in non-isotropic and relatively low-dimensional settings with $\delta\gg 0$.  

%
%
%
\subsection{Non-isotropic target and non-zero $\delta$}
\label{sec.nonisolimit}
The diffusion limit in Section \ref{seq.diffusion} was obtained as $\delta \downarrow 0$ and for an isotropic Gaussian target where the direction reflection proposals are always accepted.
In this Section, we consider the non-isotropic case of a $d$-dimensional target distribution defined as 
\begin{align}\label{eq.aniso.density}
\pi^{(d)}(x^{(d)})=\prod_{i=1}^d \gamma_i \exp\{f(\gamma_i x_i)\},
\end{align}
for inverse scalings $\gamma_i>0$ drawn independently from a distribution with second moment  $\betaE \equiv \EE(\gamma^2)$ and finite third moment. For a Discrete Bouncy Particle Sampler Markov chain exploring this density, denote by $\alpha^{(d)}_{\mathrm{pu}}\{x^{(d)},u^{(d)}\}$ and $\alpha^{(d)}_{\mathrm{dr}}\{x^{(d)},u^{(d)}\}$ the relevant acceptance probabilities in dimension $d \geq 1$. The corresponding acceptance rates are
$\alpha^{(d)}_{\mathrm{pu}}\equiv\EE[\alpha^{(d)}_{\mathrm{pu}}\{X^{(d)},U^{(d)}\}]$ and
$\alpha^{(d)}_{\mathrm{dr}}\equiv\EE[\alpha^{(d)}_{\mathrm{dr}}\{X^{(d)},U^{(d)}\}|\mathsf{DR}]$, where $\{X^{(d)},U^{(d)}\}$ follows its stationary distribution, and $\mathsf{DR}$ is the event that the initial proposal is rejected and that a {Direction Reflection} is attempted. 
The definition of the acceptance rate $\alpha^{(d)}_{\mathrm{dr}}$ requires conditioning on there being a direction update in order to ensure that the quantity $\alpha^{(d)}_{\mathrm{dr}}\{X^{(d)},U^{(d)}\}$ is well-defined. Theorem \ref{thm.noniso}, whose proof is in the \suppmat, shows that, in the high-dimensional regime $d \to \infty$ and for a fixed $\delta > 0$, the probability of accepting the direction reflection proposals converges towards one.
%
%
%
\begin{theorem}
\label{thm.noniso}Let $\{X_t^{(d)}\}_{t=0}^\infty$ be a $d$-dimensional Discrete Bouncy Particle Sampler Markov chain created by the algorithm described in Section \ref{sec.DBPS.algorithm}, exploring the target density defined in Equation \eqref{eq.aniso.density}. Assume that the function $f: \mathbb{R} \to \mathbb{R}$ has a Lipschitz second-derivative and that the quantity $J \equiv \EE\{f'(\xi)^2\}=-\EE\{f''(\xi)\}$, for a random variable $\xi$ with density $\exp\{f(\xi)\}$, is finite. The acceptance rates are such that
\begin{align*}
\lim_{d \to \infty} \; \alpha^{(d)}_{\mathrm{pu}} \, = \,  2 \, \Phi \left[-\frac{\delta}{2} \left\{\betaE \, J\right\}^{1/2} \right]
\quad \textrm{and} \qquad
\lim_{d \to \infty} \; \alpha^{(d)}_{\mathrm{dr}} \, = \, 1
\end{align*}
where $\Phi(t) = (2 \pi)^{-1/2} \, \int_0^t e^{-t^2/2} \, dt$ is the standard Gaussian cumulative function.
\end{theorem}
In the \suppmat, a simulation study verifies the theorem for a particular target distribution and a variety of time discretization parameters $\delta$. This study also suggests that, if $\delta$ is kept fixed with respect to the dimension, then $1-\alpha^{(d)}_{\mathrm{dr}}\sim 1/d$.
In Section \ref{sec.dirn.dyn}, the refreshment parameter $\kappa$ was intentionally defined so that as $\delta\downarrow 0$ the effect of $\kappa$, which is on the mixing time of the velocity refreshment, should be insensitive to the time discretization paremeter $\delta$. Given the definitions of $\kappa$ and $\delta$, it is not unreasonable to expect this insensitivity to carry through to macroscopic values of $\delta$. Section \ref{sec.sim.kappa.insens.to.delta} details a simulation study that confirms that the tuning choice for $\kappa$ is insensitive to the value of $\delta$.
This, together with Theorem \ref{thm.noniso}, suggests that the tuning strategy obtained from the isotropic Gaussian diffusion limit could be applicable for more general high-dimensional targets.

\section{Simulation studies}
\label{sec.simulations}

\subsection{Tuning of $\kappa$ is insensitive to $\delta$}
\label{sec.sim.kappa.insens.to.delta}
We first\footnote{Code is available at \href{https://github.com/alekk/discrete_bouncy_sampler}{this GitHub repository}} investigate the robustness to the choice of $\delta$ of the effect of $\kappa$ on efficiency. 
We considered an isotropic Gaussian target distribution with dimension $d=100$ and, for a range of $\delta\in\{0.04,0.2,1.0\}$, we estimated the efficiency of the \DBPS~as a function of the refreshment parameter $\kappa$. 
%
%
For a fixed $\delta$, the relative variation in $1-\alpha_{\mathrm{pu}}$ was small, varying less than $5\%$ from a central value over the whole range of $\kappa$ values. Central values of $1-\alpha_{\mathrm{pu}}$ were $2\%$ ($\delta=0.04$),  $8 \%$ ($\delta=0.2$) and $38 \%$ ($\delta=1.0$). These values span the range of potential interest in the applications we have looked at.
\begin{figure}[t]
\begin{center}
\includegraphics[scale=0.5]{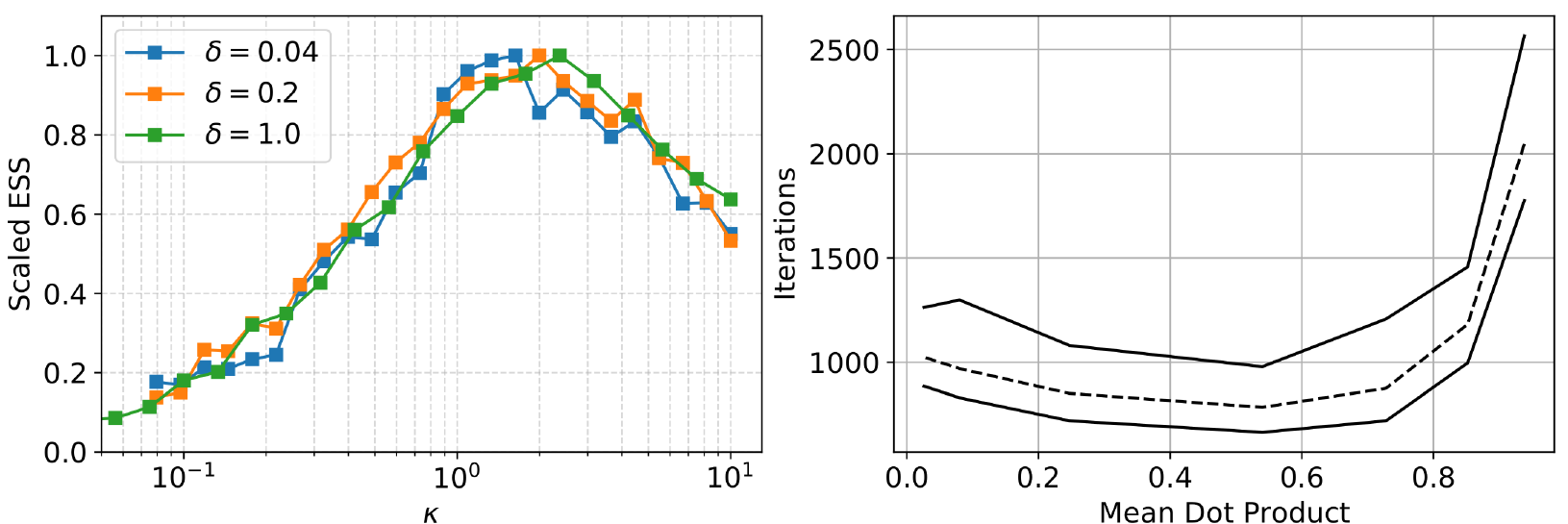}
\end{center}
\caption{Left: relative effective sample size against $\kappa$ for a $100$-dimensional standard Gaussian target. Right: maximum (solid), minimum (solid) and median (dashed) iterations to converge from a random tail point of target \eqref{eqn.powaniso} as a function of the mean dot product statistic $\beta$ in a stationary run.}
\label{fig.simstud.optkapinsendel}
\end{figure}
The left panel of Figure \ref{fig.simstud.optkapinsendel} suggests that the optimal value of the refreshment parameter $\kappa$ is insensitive to the time discretization parameter $\delta$.
This insensitivity over more than an order of magnitude of $\delta$ values suggests that the value of the optimal refreshment parameter $\kappa$ obtained in the limit $\delta \to 0$ and, hence, $1-\alpha_{\mathrm{pu}}\rightarrow 0$, is indicative of the optimal refreshment parameter $\kappa$ when $\delta$ is macroscopic and $1-\alpha_{\mathrm{pu}} \gg 0$.

\subsection{Robustness of advice to departures from Gaussianity and isotropy}
\label{sec.sim.simple}
We next investigate the robustness of our tuning advice to departures of the target from Gaussianity and isotropy.

\begin{figure}[t]
\begin{center}
\includegraphics[scale=0.6]{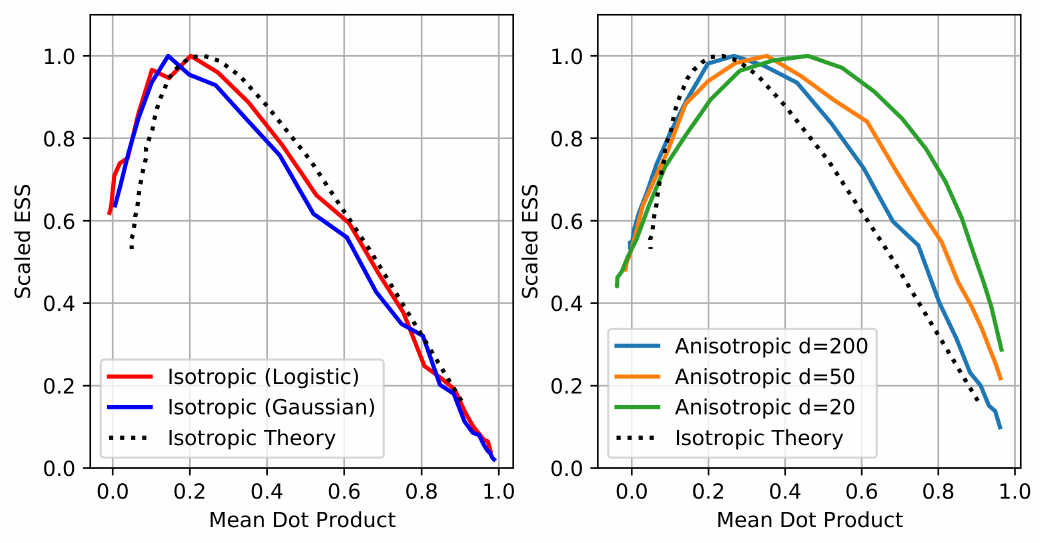}
\end{center}
\caption{Scaled effective sample size as a function of the mean dot product for several values of $\kappa$, obtained from MCMC runs of length $N=10^6$. The \DBPS is applied to the target distributions $\pi^{\textrm{log}}$ and $\pi^{\textrm{iso}}$ (Left) and $\pi^{\textrm{aniso}}$ using $d=20$, $d=50$ and $d=200$ (Right)}
\label{fig.toy.ess}
\end{figure}
Consider three scenarios: an isotropic multivariate logistic distribution $\pi^{\textrm{logis}}$ with density\\ $\prod_{i=1}^d \, \exp(x_i) / [1+\exp(x_i)]^2$, an isotropic Gaussian distribution $\pi^{\textrm{iso}}$ with density proportional to $\prod_{i=1}^d \, \exp(-x_i^2/2)$ and a non-isotropic multivariate Gaussian distribution $\pi^{\textrm{aniso}}$ with density proportional to $\prod_{i=1}^d \, \exp\{-x_i^2/(2 \, \sigma_i^2) \}$. In this section, we choose $d=100$ for both isotropic targets, and $d \in \{20, 50, 200\}$ for the anisotropic target. In order to test the robustness of our tuning guidelines to non-isotropic distributions, we chose the scales $\sigma_1 < \ldots < \sigma_d$ linearly separated between $\sigma_1=1$ and $\sigma_d=10$. The scaled effective sample size  curves as a function of the mean dot-product are displayed in Figure \ref{fig.toy.ess}. There is extremely good agreement with the theory developed in Section \ref{sec.simulations} for the isotropic distribution $\pi^{\textrm{iso}}$ and the approximately isotropic distribution $\pi^{\textrm{logistic}}$. Not surprisingly, mild departure from the theory is observed for strongly non-isotropic distributions such as $\pi^{\textrm{aniso}}$. However, for dimension $d=200$ the departure, especially in terms of the optimal dot product, is barely noticeable. When $d=20$ and $d=50$, however, our proposed guideline, \emph{i.e.} tune the refreshment rate $\kappa$ such that the mean dot-product $\beta \approx 0.2$, leads only to a loss of efficiency of approximately $10\%$  and $5\%$ respectively.

\subsection{Convergence and tail behaviour}
One of the most commonly used algorithms for inference in high-dimensional scenarios is Hamiltonian Monte Carlo \cite[][]{duane1987hybrid}. However, it is well known \citep{livingstone2019} that due to the dependence of the Leapfrog step on $\|\nabla \log \pi\|$, Hamiltonian Monte Carlo is not geometrically ergodic on targets with tails lighter than those of a Gaussian. By contrast, the \DBPS  depends on  $\nabla \log \pi$ only through the equivalent normalized vector. The \suppmat~ details a simulation study on a non-isotropic, light-tailed target where a tuned Hamiltonian Monte Carlo algorithm is nearly $50\%$ more efficient than a tuned discrete bouncy particle sampler when started from stationarity. However, with the same tunings, but when started from a random point in the tail of the distribution, Hamiltonian Monte Carlo does not even move, whereas the discrete bouncy particle sampler quickly converges to the centre of the posterior mass. 

Whilst our theory combined with empirical verifications suggests that the optimal refreshment rate $\kappa$ at stationarity is that which leads to an average dot product $\beta \approx 0.2$, it may be that a different value is optimal for convergence from the tails. Finally, therefore, we examine the effect of $\kappa$, when the \DBPS is used to explore the target distribution in $\mathbb{R}^d$ with $d=50$ and a density of
\begin{align}\label{eqn.powaniso}
f(x) &\propto \exp\BK{ -\frac{1}{a} \|x\|^a_{\text{M}} }
\qquad \textrm{for} \quad
\|x\|^2_{\text{M}} \equiv \sum_{i=1}^d \frac{x_i^2}{\sigma_i^2}
\end{align}
where $a=4$ and $\sigma_1,\dots,\sigma_d$ are as described in Section \ref{sec.sim.simple}. In $d>1$ dimensions, the modal value for $\|x\|_{\text{M}}$ is $r_\star \equiv (d-1)^{1/a}$. 
For several values of $\kappa > 0$, we repeated the following experiment $n=100$ times: sample a random unit vector $z\in\mathbb{S}^{d-1}$, set the initial position $x_0=10 \, r_\star \cdot (\sigma_1 \, z_1, \dots, \sigma_d z_d) \in \mathbb{R}^d$ far out in the tail of the distribution, run the \DBPS with $\delta=2.0$ (a sensible value to explore the body of the target) until $\| x\|_{\text{M}} \leq r_\star$ and note the iteration number at which that happened. 
The right-hand panel of Figure \ref{fig.simstud.optkapinsendel} shows the minimum, maximum and median iteration number at which convergence (by this measure) was achieved. Except for very low $\kappa$ values, which lead to large dot product statistics, the behaviour is robust to the choice of $\kappa$, suggesting that it is reasonable to apply our tuning criterion when the chain is started away from its stationary distribution.

%
%
\subsection{The Markov modulated Poisson process}
\label{sec.mmpp}
A simulation study on the eight-dimensional posterior of a non-trivial statistical model, the Markov modulated Poisson process, is detailed in the \suppmat. We find that mixing efficiency of $\log \pi$ is optimized at a dot product statistic of $\beta \approx 0.4$, tuning to $\beta \approx 0.2$ would lead to only a $10\%$ reduction in efficiency. When either preconditioning or using only $n_{\text{cpt}}=3$ random components of the gradient vector, the optimal efficiency is achieved for a dot product statistics of $\beta \approx 0.2$.

\section{Discussion}
The key advantage of the Discrete Bouncy Particle Sampler over its continuous-time counterpart is that posterior and gradient evaluations can be treated as a ``black box'' with no requirement to bound the gradient so as to apply Poisson thinning. Unlike the continuous-time algorithm, there is a chance that an attempted bounce will be rejected and the particle will approximately backtrack, however for sensible tunings, the back-tracking probability converges towards zero as the dimension of the problem increases. 

The average computational cost per iteration is insensitive to the choice of $\kappa$ since the direction is updated every iteration, and $\kappa$ has little effect on the number of steps between potential bounces. Thus it is sufficient for the purposes of this article to describe and empirically record efficiency in terms of effective sample size rather than effective samples per unit of time. The only exception to this is when we compare against Hamiltonian Monte Carlo in Appendix \ref{sec.convfromtail}.

The dot-product tuning diagnostic maps $\kappa$ to an absolute scale via the properties of the posterior, just as the acceptance rate diagnostic does for the scaling in the random-walk Metropolis algorithm. When $\kappa=0$ the velocity direction just before the next bounce is identical to that just after the previous bounce, and the dot product is unity. The mixing time of the refreshment process is $1/\kappa$; when this is small compared with the time between bounces or, equivalently, the length scale of the target, the two velocity directions bear little relation to each other, and the dot product is small. 

Whilst this article offers theory-based practical advice on the tuning of the refreshment parameter, $\kappa$, it does not tackle the choice of the discretization parameter, $\delta$. In contrast to the insensitivity of computational cost to the choice of $\kappa$, increasing $\delta$ increases the frequency of potential bounces and hence of expensive gradient calculations, and this would need to be accounted for in any analysis.

\section*{Acknowledgements}
Work by CS was supported by EPSRC grant EP/P033075/1. AHT acknowledges support from the Singapore Ministry of Education Tier 2 Grant  (MOE2016-T2-2-135) and a Young Investigator Award Grant (NUSYIA FY16 P16; R-155-000-180-133).

\bibliographystyle{natbib}
\bibliography{references}

\begin{thebibliography}{37}
\expandafter\ifx\csname natexlab\endcsname\relax\def\natexlab#1{#1}\fi

\bibitem[{Andrieu et~al.(2018)Andrieu, Durmus, N{\"u}sken \&
  Roussel}]{andrieu2018hypercoercivity}
\textsc{Andrieu, C.}, \textsc{Durmus, A.}, \textsc{N{\"u}sken, N.} \&
  \textsc{Roussel, J.} (2018).
\newblock Hypercoercivity of piecewise deterministic {M}arkov process-{M}onte
  {C}arlo.
\newblock \textit{arXiv preprint arXiv:1808.08592} .

\bibitem[{Beskos et~al.(2013)Beskos, Pillai, Roberts, Sanz-Serna, Stuart
  et~al.}]{beskos2013optimal}
\textsc{Beskos, A.}, \textsc{Pillai, N.}, \textsc{Roberts, G.},
  \textsc{Sanz-Serna, J.-M.}, \textsc{Stuart, A.} et~al. (2013).
\newblock Optimal tuning of the hybrid {M}onte {C}arlo algorithm.
\newblock \textit{Bernoulli} \textbf{19}, 1501--1534.

\bibitem[{Betancourt et~al.(2014)Betancourt, Byrne \&
  Girolami}]{betancourt2014optimizing}
\textsc{Betancourt, M.}, \textsc{Byrne, S.} \& \textsc{Girolami, M.} (2014).
\newblock Optimizing the integrator step size for {H}amiltonian {M}onte
  {C}arlo.
\newblock \textit{arXiv preprint arXiv:1411.6669} .

\bibitem[{Bierkens et~al.(2019)Bierkens, Fearnhead \&
  Roberts}]{bierkens2019zig}
\textsc{Bierkens, J.}, \textsc{Fearnhead, P.} \& \textsc{Roberts, G.} (2019).
\newblock The zig-zag process and super-efficient sampling for {B}ayesian
  analysis of big data.
\newblock \textit{Ann. Statist.} \textbf{47}, 1288--1320.

\bibitem[{Bierkens et~al.(2018)Bierkens, Kamatani \&
  Roberts}]{bierkens2018high}
\textsc{Bierkens, J.}, \textsc{Kamatani, K.} \& \textsc{Roberts, G.~O.} (2018).
\newblock High-dimensional scaling limits of piecewise deterministic sampling
  algorithms.
\newblock \textit{arXiv preprint arXiv:1807.11358} .

\bibitem[{Bierkens \& Roberts(2017)}]{bierkens2017piecewise}
\textsc{Bierkens, J.} \& \textsc{Roberts, G.} (2017).
\newblock A piecewise deterministic scaling limit of lifted
  {M}etropolis--{H}astings in the {C}urie--{W}eiss model.
\newblock \textit{The Annals of Applied Probability} \textbf{27}, 846--882.

\bibitem[{Bouchard-C{\^o}t{\'e} et~al.(2017)Bouchard-C{\^o}t{\'e}, Vollmer \&
  Doucet}]{bouchard2017bouncy}
\textsc{Bouchard-C{\^o}t{\'e}, A.}, \textsc{Vollmer, S.~J.} \& \textsc{Doucet,
  A.} (2017).
\newblock The bouncy particle sampler: A non-reversible rejection-free {M}arkov
  chain {M}onte {C}arlo method.
\newblock \textit{Journal of the American Statistical Association} .

\bibitem[{Chen et~al.(1999)Chen, Lov{\'a}sz \& Pak}]{chen1999lifting}
\textsc{Chen, F.}, \textsc{Lov{\'a}sz, L.} \& \textsc{Pak, I.} (1999).
\newblock Lifting {M}arkov chains to speed up mixing.
\newblock In \textit{Proceedings of the thirty-first annual ACM symposium on
  Theory of computing}. ACM.

\bibitem[{Deligiannidis et~al.(2021)Deligiannidis, Paulin \&
  Doucet}]{deligiannidis2018randomized}
\textsc{Deligiannidis, G.}, \textsc{Paulin, D.} \& \textsc{Doucet, A.} (2021).
\newblock Randomized {H}amiltonian {M}onte {C}arlo as scaling limit of the
  bouncy particle sampler and dimension-free convergence rates.
\newblock \textit{Annals of Applied Probability} Accepted.

\bibitem[{Diaconis et~al.(2000)Diaconis, Holmes \& Neal}]{diaconis2000analysis}
\textsc{Diaconis, P.}, \textsc{Holmes, S.} \& \textsc{Neal, R.~M.} (2000).
\newblock Analysis of a nonreversible {M}arkov chain sampler.
\newblock \textit{Annals of Applied Probability} , 726--752.

\bibitem[{Duane et~al.(1987)Duane, Kennedy, Pendleton \&
  Roweth}]{duane1987hybrid}
\textsc{Duane, S.}, \textsc{Kennedy, A.~D.}, \textsc{Pendleton, B.~J.} \&
  \textsc{Roweth, D.} (1987).
\newblock Hybrid {M}onte {C}arlo.
\newblock \textit{Physics letters B} \textbf{195}, 216--222.

\bibitem[{Duncan et~al.(2016)Duncan, Lelievre \&
  Pavliotis}]{duncan2016variance}
\textsc{Duncan, A.~B.}, \textsc{Lelievre, T.} \& \textsc{Pavliotis, G.} (2016).
\newblock Variance reduction using nonreversible {L}angevin samplers.
\newblock \textit{Journal of statistical physics} \textbf{163}, 457--491.

\bibitem[{Fearnhead et~al.(2018)Fearnhead, Bierkens, Pollock \&
  Roberts}]{FBPR2018}
\textsc{Fearnhead, P.}, \textsc{Bierkens, J.}, \textsc{Pollock, M.} \&
  \textsc{Roberts, G.~O.} (2018).
\newblock Piecewise-deterministic {M}arkov processes for continuous-time
  {M}onte {C}arlo.
\newblock \textit{Statist. Sci.} \textbf{33}, 386--412.

\bibitem[{Fearnhead \& Sherlock(2006)}]{FearnheadSherlock2006}
\textsc{Fearnhead, P.} \& \textsc{Sherlock, C.} (2006).
\newblock An exact {G}ibbs sampler for the {M}arkov-modulated {P}oisson
  process.
\newblock \textit{J. R. Stat. Soc. Ser. B Stat. Methodol.} \textbf{68},
  767--784.

\bibitem[{Fielding et~al.(2011)Fielding, Nott \& Liong}]{fielding2011efficient}
\textsc{Fielding, M.}, \textsc{Nott, D.~J.} \& \textsc{Liong, S.-Y.} (2011).
\newblock Efficient {MCMC} schemes for computationally expensive posterior
  distributions.
\newblock \textit{Technometrics} \textbf{53}, 16--28.

\bibitem[{Gustafson(1998)}]{gustafson1998guided}
\textsc{Gustafson, P.} (1998).
\newblock A guided walk {M}etropolis algorithm.
\newblock \textit{Statistics and Computing} \textbf{8}, 357--364.

\bibitem[{Hwang et~al.(2015)Hwang, Normand \& Wu}]{hwang2015variance}
\textsc{Hwang, C.-R.}, \textsc{Normand, R.} \& \textsc{Wu, S.-J.} (2015).
\newblock Variance reduction for diffusions.
\newblock \textit{Stochastic Processes and their Applications} \textbf{125},
  3522--3540.

\bibitem[{Leli{\`e}vre et~al.(2013)Leli{\`e}vre, Nier \&
  Pavliotis}]{lelievre2013optimal}
\textsc{Leli{\`e}vre, T.}, \textsc{Nier, F.} \& \textsc{Pavliotis, G.~A.}
  (2013).
\newblock Optimal non-reversible linear drift for the convergence to
  equilibrium of a diffusion.
\newblock \textit{Journal of Statistical Physics} \textbf{152}, 237--274.

\bibitem[{Livingstone et~al.(2019)Livingstone, Betancourt, Byrne \&
  Girolami}]{livingstone2019}
\textsc{Livingstone, S.}, \textsc{Betancourt, M.}, \textsc{Byrne, S.} \&
  \textsc{Girolami, M.} (2019).
\newblock On the geometric ergodicity of {H}amiltonian {M}onte {C}arlo.
\newblock \textit{Bernoulli} \textbf{25}, 3109--3138.

\bibitem[{Michel et~al.(2020)Michel, Durmus \&
  S{\'e}n{\'e}cal}]{michel2020forward}
\textsc{Michel, M.}, \textsc{Durmus, A.} \& \textsc{S{\'e}n{\'e}cal, S.}
  (2020).
\newblock Forward event-chain {M}onte {C}arlo: Fast sampling by randomness
  control in irreversible {M}arkov chains.
\newblock \textit{Journal of Computational and Graphical Statistics} , 1--14.

\bibitem[{Monmarch{\'e}(2019)}]{monmarche2019kinetic}
\textsc{Monmarch{\'e}, P.} (2019).
\newblock Kinetic walks for sampling.
\newblock \textit{arXiv preprint arXiv:1903.00550} .

\bibitem[{Neal(2003)}]{neal2003}
\textsc{Neal, R.~M.} (2003).
\newblock Slice sampling.
\newblock \textit{Ann. Statist.} \textbf{31}, 705--767.

\bibitem[{Pakman et~al.(2017)Pakman, Gilboa, Carlson \&
  Paninski}]{Pakmanetal2017}
\textsc{Pakman, A.}, \textsc{Gilboa, D.}, \textsc{Carlson, D.} \&
  \textsc{Paninski, L.} (2017).
\newblock Stochastic bouncy particle sampler.
\newblock In \textit{Proceedings of the 34th International Conference on
  Machine Learning}, D.~Precup \& Y.~W. Teh, eds., vol.~70 of
  \textit{Proceedings of Machine Learning Research}. International Convention
  Centre, Sydney, Australia: PMLR.

\bibitem[{Papanicolaou(1977)}]{papanicolaou1977martingale}
\textsc{Papanicolaou, G.} (1977).
\newblock Martingale approach to some limit theorems.
\newblock In \textit{Papers from the Duke Turbulence Conference, Duke Univ.,
  Durham, NC, 1977}.

\bibitem[{Pavliotis \& Stuart(2008)}]{pavliotis2008multiscale}
\textsc{Pavliotis, G.} \& \textsc{Stuart, A.} (2008).
\newblock \textit{Multiscale methods: averaging and homogenization}.
\newblock Springer Science \& Business Media.

\bibitem[{Peters \& de~With(2012)}]{peters2012rejection}
\textsc{Peters, E.~A.} \& \textsc{de~With, G.} (2012).
\newblock Rejection-free {M}onte {C}arlo sampling for general potentials.
\newblock \textit{Physical Review E} \textbf{85}, 026703.

\bibitem[{Ramm \& Smirnova(2001)}]{ramm2001stable}
\textsc{Ramm, A.} \& \textsc{Smirnova, A.} (2001).
\newblock On stable numerical differentiation.
\newblock \textit{Mathematics of computation} \textbf{70}, 1131--1153.

\bibitem[{Rey-Bellet \& Spiliopoulos(2015)}]{rey2015irreversible}
\textsc{Rey-Bellet, L.} \& \textsc{Spiliopoulos, K.} (2015).
\newblock Irreversible {L}angevin samplers and variance reduction: a large
  deviations approach.
\newblock \textit{Nonlinearity} \textbf{28}, 2081.

\bibitem[{Roberts et~al.(1997)Roberts, Gelman, Gilks et~al.}]{roberts1997weak}
\textsc{Roberts, G.~O.}, \textsc{Gelman, A.}, \textsc{Gilks, W.~R.} et~al.
  (1997).
\newblock Weak convergence and optimal scaling of random walk {M}etropolis
  algorithms.
\newblock \textit{The annals of applied probability} \textbf{7}, 110--120.

\bibitem[{Roberts \& Rosenthal(2001)}]{RobRos2001}
\textsc{Roberts, G.~O.} \& \textsc{Rosenthal, J.~S.} (2001).
\newblock Optimal scaling for various {M}etropolis-{H}astings algorithms.
\newblock \textit{Statist. Sci.} \textbf{16}, 351--367.

\bibitem[{Roberts \& Rosenthal(2016)}]{roberts2016complexity}
\textsc{Roberts, G.~O.} \& \textsc{Rosenthal, J.~S.} (2016).
\newblock Complexity bounds for {M}arkov chain {M}onte {C}arlo algorithms via
  diffusion limits.
\newblock \textit{Journal of Applied Probability} \textbf{53}, 410--420.

\bibitem[{Terenin \& Thorngren(2018)}]{terenin2018piecewise}
\textsc{Terenin, A.} \& \textsc{Thorngren, D.} (2018).
\newblock A piecewise deterministic {M}arkov process via radius-angle swaps in
  hyperspherical coordinates.
\newblock \textit{arXiv preprint arXiv:1807.00420} .

\bibitem[{Tierney \& Mira(1999)}]{TierneyMira1999}
\textsc{Tierney, L.} \& \textsc{Mira, A.} (1999).
\newblock Some adaptive {M}onte {C}arlo methods for {B}ayesian inference.
\newblock \textit{Statistics in Medicine} \textbf{18}, 2507--2515.

\bibitem[{Vanetti et~al.(2017)Vanetti, Bouchard-C{\^o}t{\'e}, Deligiannidis \&
  Doucet}]{VanettiBouchardDelDoucet2017}
\textsc{Vanetti, P.}, \textsc{Bouchard-C{\^o}t{\'e}, A.},
  \textsc{Deligiannidis, G.} \& \textsc{Doucet, A.} (2017).
\newblock Piecewise-deterministic {M}arkov chain {M}onte {C}arlo.
\newblock \textit{arXiv preprint arXiv:1707.05296} .

\bibitem[{Weinan(2011)}]{weinan2011principles}
\textsc{Weinan, E.} (2011).
\newblock \textit{Principles of multiscale modeling}.
\newblock Cambridge University Press.

\bibitem[{{Wu} \& {Robert}(2017)}]{WuRobert2017}
\textsc{{Wu}, C.} \& \textsc{{Robert}, C.~P.} (2017).
\newblock {Generalized Bouncy Particle Sampler}.
\newblock \textit{ArXiv e-prints} .

\bibitem[{Wu \& Robert(2020)}]{wu2018coordinate}
\textsc{Wu, C.} \& \textsc{Robert, C.~P.} (2020).
\newblock The coordinate sampler: A non-reversible {G}ibbs-like {MCMC} sampler.
\newblock \textit{Statistics and Computing} \textbf{30}, 721--730.

\end{thebibliography}

\appendix

\section{Correctness and proofs of propositions}
\setlist[description]{font=\normalfont\itshape}

%
%
\subsection{Correctness of the Discrete Bouncy Particle Sampler}
In this section, we prove the correctness of a slightly more general version of the Discrete Bouncy Particle Sampler than the one described in the main text. This added generality is exploited in Section \ref{sec.approxgrad}.
Consider a generalized reflection operator $\mathsf{R}: \bR^d \times \bR^d$ such that for every $x \in \cX \equiv \bR^d$, the mapping $u \mapsto \mathsf{R}(u,x)$ satisfies the following three conditions:
\begin{description}
\item[B1] For any $u \in \bR^d$,  we have that $\mathsf{R}(-\mathsf{R}(u,x), x) = -u$
\item[B2] The mapping $u \mapsto \mathsf{R}(u,x)$ is volume preserving. 
\item[B3]The mapping $u \mapsto \mathsf{R}(u,x)$ preserves norms, $\|\mathsf{R}(u,x)\| = \|u\|$.
\end{description}
For $(x_{k}, u_{k}) \in \bR^d \times \bR^d$ and a time discretization parameter $\delta > 0$, consider the Markov kernel $(x_{k}, u_{k}) \mapsto (x_{k+1}, u_{k+1})$ defined as the composition of the following three steps. 
\begin{description}
\item[Step 1.] Generate a proposal $(x',u')=(x_k+ \delta \, u_k, -u_k)$. With probability
\begin{align*}
\alpha_{\mathrm{pu}}(x_k, u_k)\equiv 1 \, \wedge \, \frac{ \pi(x') }{ \pi(x_k) },
\end{align*}
set $(\hat{x}_{k},\hat{u}_{k})=(x',-u')$ and go to { Step 3.} Otherwise, proceed to { Step 2.}
\item[Step 2.] 
consider $u''= -\mathsf{R}(u, x')$ and $x'' = x' - \delta \,u''$. With probability
%
%
\begin{align*}
\alpha_{\mathsf{R}}(x_k, u_k)\equiv 1 \, \wedge \, 
\left\{
\frac{ 1-\alpha_{\mathrm{pu}}(x'',u'') }{1-\alpha_{\mathrm{pu}}(x_k,u_k)} \times \frac{\pi(x'')}{\pi(x_k)} \right\},
\end{align*}
set $(\hat{x}_{k},\hat{u}_{k})=(x'',u'')$. Otherwise, set $(\hat{x}_{k},\hat{u}_{k})=(x_k,u_k)$.

\item[Step 3.] Reverse the direction: $(x_{k+1}, u_{k+1}) = (\hat{x}_{k}, -\hat{u}_{k})$
\end{description}

\begin{lemma} \label{lem.correct} Consider any spherically symmetric probability density $\rho(u)$. Under Assumptions {B1-2-3}, the Markov kernel described by {Step 1-2-3} leaves the density $\widetilde{\pi}(x,u) = \pi(x) \, \rho(u)$ invariant.
\end{lemma}
%
%
\begin{proof}
Since $\rho$ is spherically symmetric and $\mathsf{R}$ preserves norms, { Step 3} leaves $\widetilde{\pi}$ invariant. Now, the combination of {Step 1-2} is exactly a {Delayed Rejection} \citep{TierneyMira1999} Markov kernel with two proposal mechanisms: $(x,u) \mapsto (x + \delta \, u, -u)$ and $(x,u) \mapsto (x + \delta \, u  + \delta \, \mathsf{R}(u, x + \delta u), - \mathsf{R}(u, x + \delta u))$ and target density $\pi(x) \, \rho(x)$. Algebra directly shows that these two proposals are volume preserving involutions. To conclude the proof of the lemma, it thus suffices to show that the usual delayed rejection scheme for sampling from a density $\mu(z)$ on a state-space $\mathcal{Z} \subset \bR^N$ remains valid with deterministic proposals $z \mapsto z' = T_1(z)$ and $z \mapsto z''=T_2(z)$ that are volume preserving involutions. For an arbitrary bounded test function $\phi: \mathcal{Z} \to \bR$, one needs to show that
\begin{align*}
\int 
\left[ \phi\left\{T_1(z)\right\} \, \alpha_1(z)+
\phi\left\{T_2(z)\right\} \, \left\{1-\alpha_1(z)\right\} \, \alpha_2(z)+ 
\phi(z) \, \alpha_3(z) \right] \, \mu(z) \, \md z 
= \int \phi(z)  \, \mu(z) \, \md z
\end{align*}
with $\alpha_1(z)= 1 \wedge \mu\{T_1(z)\} / \mu(z)$ and $\alpha_3(z)= 1 - \alpha_1(z) - \{1-\alpha_1(z)\} \, \alpha_2(z)$ and
\begin{align*}
\alpha_2(z) = 1 \, \wedge \, 
\curBK{
\frac{ 1-\alpha_1\{T_2(z)\} }{1-\alpha_1(z)} \times \frac{\mu\{T_2(z)\}}{\mu(z)}}.
\end{align*}
Algebra shows that this is equivalent to proving that
\begin{align} \label{eq.DR.proof.bis}
\begin{aligned}
\int 
&\sqBK{ \phi\{T_1(z)\} - \phi(z) } \times \sqBK{\mu(z) \wedge \mu\{T_1(z)\} }\\
&\; 
+ \int
\sqBK{ \phi\{T_2(z)\} - \phi(z) } \times \sqBK{\mu(z)\{1-\alpha_1(z)\} \wedge \mu\{T_2(z)\}(1-\alpha_1\{T_2(z)\}) }  \, \md z 
= 0.
\end{aligned}
\end{align}
Since $T_1$ and $T_2$ are involutions that preserve volume, a change of variable $z \mapsto T_1(z)$ shows that the first integral in Equation \eqref{eq.DR.proof.bis} also equals its negation, and hence vanishes. And similarly, the change of variable $z \mapsto T_2(z)$ shows that the second integral in Equation \eqref{eq.DR.proof.bis} also vanishes. This concludes the proof of the lemma.
\end{proof}
In Section \ref{sec.DBPS.algorithm}, the combination of the Position Update and  Direction Update is equivalent to {Step 1-2-3} with the operator $\mathsf{R}(u,x) = \reflect_{\field(x)}(u)$. Since algebra shows that the Conditions {B1-2-3} are satisfied, Lemma \ref{lem.correct} thus shows the correctness of the Discrete Bouncy Particle Sampler as described in Section \ref{sec.DBPS.algorithm}.\\

Indeed, one can consider mixtures of operators that satisfy Conditions {B1-2-3}. Namely, for a conditional probability distribution $\mathrm{M}(x, \md \omega)$ and an operator $\widetilde{\mathsf{R}}: \bR^d \times \bR^d \times \Omega \to \bR^d$ such that for any value of $(x,\omega) \in \bR^d \times \Omega$ the operator $u \mapsto \widetilde{\mathsf{R}}(u,x,\omega)$ satisfies the Conditions {B1-2-3},  one can consider the Markov kernel that, for a given pair $(x,u) \in \bR^d \times \bR^d$, starts by generating $\omega \sim \mathrm{M}(x, \md \omega)$ and then proceeds to {Steps 1-2-3} with generalized operator $u \mapsto \widetilde{\mathsf{R}}(u,x,\omega)$. Lemma \ref{lem.correct} shows that this leads to a valid algorithm. This in turns shows that, in Section \ref{sec.DBPS.algorithm}, one can also consider randomized vector fields: instead of performing a reflection with respect to a fixed vector field $\field(x)$,  one can instead generate a random vector $v \in \cX \setminus \{0\}$ sampled from a distribution that depends on the vector $x \in \cX$ only (i.e. does not depend on the current direction $u \in \cS$) and attempts the reflection $u \mapsto \reflect_v(u)$. This remark also shows the correctness of the methods described in Section \ref{sec.approxgrad}.

%
%
\subsection{Proof of Proposition \ref{prop.delta.zero.limit}}
Recall that the quantity $\lambda(x,u)$ is defined as $\lambda(x,u) \equiv \bra{-\nabla \log \pi(x), u}_+$. Under Assumption {(A1)} and a discretization parameter $\delta > 0$, the acceptance probability $\alpha^{\delta}(x,u)$that the proposal $(x,u) \mapsto (x +\delta \, u, u)$ is accepted reads
\begin{align}\label{eq.alpha.expansion}
\begin{aligned}
\alpha^{\delta}(x,u) 
&= \exp \sqBK{ \min\curBK{0, \log \pi(x+\delta \,u) - \log \pi(x,u)}} \\
&= \exp \left\{ \delta \times \bra{\nabla \log \pi(x), u}_-\right\}  \; + \; \mathcal{O}(\delta^2) \\
&= \exp \left\{ - \delta \times \lambda(x,u) \right\}  \; + \; \mathcal{O}(\delta^2)
\end{aligned}
\end{align}
where $\mathcal{O}(\delta^2)$ is a quantity whose absolute value is less than a constant times $\delta^2$.
For $t > 0$, the probability that the \DBPS algorithm accepts $\floor{t / \delta}+1$ consecutive proposals $(x,u) \mapsto (x+\delta \,u, u)$ without reflection attenpts equals $\prod_{k=0}^{\floor{t / \delta}} \alpha^{\delta}(x^{\delta}_k, u^{\delta}_k)$. Under Assumption {(A3)}, one can condition upon a fixed trajectory of the Markov process $V$, i.e. $V_t = v_t$ for all $0 \leq t \leq T$ and $u^{\delta}_k = \overline{u}^{\delta}_{k \delta} = v_{k \delta}$, not depending on the parameter $\delta$, so that $\overline{x}^{\delta}_{k \delta} = \overline{x}^{\delta}_{0} + \delta \sum_{j=0}^{k-1} v_{j \delta}$.
Equation \eqref{eq.alpha.expansion}, the continuity of the rate function $\lambda$ as well as the continuity of the trajectories of the Markov process $V$, show that 
\begin{align*}
\lim_{\delta \to 0} \, \prod_{k=0}^{\floor{t / \delta}} \alpha^{\delta}(x^{\delta}_k, u^{\delta}_k)
&=
\lim_{\delta \to 0} \, \exp \left\{ - \delta \, \sum_{k=0}^{\floor{t / \delta}} \lambda(\overline{x}^{\delta}_{k \delta} , v_{k \delta})\right\} + \mathcal{O}(\delta)
=
\exp \left\{ -\int_{0}^t \lambda(\overline{x}_s, v_s) \, \md s \right\}
\end{align*}
where $\overline{x}_s = x_0 + \int_0^s v_t \, dt$.
This means that, in the limit $\delta \to 0$, bounce attempts arrive 
at rate $\lambda(x,u)$ and, in between the bounces, the limiting process simply evolves according to the dynamics \eqref{eq.X.dynamics}.

Finally, once a proposal $(x,u) \mapsto (x+\delta \,u, u)$ is rejected, the second proposal $(x,u) \mapsto (x'', u'')$, i.e. the bounce, is accepted with 
probability $\alpha_{\textrm{DR}}(x,u)$ described in Equation \eqref{eq.alpha.DR}. 
By continuity of the density $x \mapsto \pi(x)$, we have that $\pi(x'') / \pi(x) \to 1$ as $\delta \to 0$. Furthermore, the Taylor expansion \eqref{eq.alpha.expansion} gives that $1-\alpha^\delta(x,u) = \delta \, \lambda(x,u) + \mathcal{O}(\delta^2)$. Under Assumption, the vector field $\field$ is continuous, which implies that
\begin{align} \label{eq.DR.expansion}
\lim_{\delta \to 0} \;
\frac{ 1-\alpha^\delta(x'',-u'') }{1-\alpha^\delta(x,u)} \times \frac{\pi(x'')}{\pi(x)}
\;=\;
\frac{\lambda\{x, -\reflect_{\field(x)}(u)\}}{\lambda(x,u)},
\end{align}
where we have dropped the dependence on $\delta$ from the notation $x'' = x + \delta \, u + \delta \, \reflect_{\field(x + \delta u)}(u)$ and $u'' = \reflect_{\field(x + \delta u)}(u)$.
It follows from \eqref{eq.DR.expansion} that, in the limit as $\delta \to 0$, a proposed bounce $(x,u) \mapsto (x'',u'')$ is accepted with probability $\mathcal{A}(x,u)$ described in Equation \eqref{eq.limiting.acceptance}, with $(x'',u'') \to (x, \reflect_{\field(x)}(u))$ as $\delta \to 0$. This completes the proof of Proposition \ref{prop.delta.zero.limit}.

\subsection{Proof of Proposition \ref{prop.diffusion.limit}}
For a fixed dimension $d \geq 1$, Proposition \ref{prop.delta.zero.limit} describes its scaling limit as $\delta \to 0$: the processes $t \mapsto z^{d,\delta}_t = (x^{d,\delta}_{\delta t}, u^{d,\delta}_{\delta t})$ converges on path-space to the jump diffusion that evolves according to $d \XX^d_t = \UU^d_t \, dt$, where $\UU^d_t$ is a Brownian motion on the unit sphere of $\mathbb{R}^d$ whose dynamics is described by the Stochastic Differential Equation \eqref{eq.brownian.SDE}, in between reflections $\UU^d_t = \reflect\BK{ \UU^d_{t^-}, \nabla \log \pi_d(\XX^d_{t^-}) }$ that arrive at rate $\lambda(\XX^d_t, \UU^d_t) = \bra{-\nabla \pi_d( \XX^d_t), \UU^d_t}_+$. The generator of the joint process $(\XX^d_t, \UU^d_t)$ reads
\begin{align*}
\phi(x,u) \mapsto \bra{u ,  \nabla_x \phi(x,u)} + \cL^{(\kappa,B)} \phi(x,u) + \lambda(x,u) \, \cF \phi(x,u)
\end{align*}
where $\cL^{(\kappa,B)}$ is the generator of the Brownian motion on the united sphere \eqref{eq.brownian.SDE} and $\cF$ is the flip operator defined as $\cF \phi(x,u) = \phi(x,-u) - \phi(x,u)$. In order to obtain the limit of the process defined in Equation \eqref{eq.radial.process}, set
\begin{align*}
R^d_t \equiv \| \XX^d_{d \times t}\| - \sigma \, d^{1/2}
\qquad \textrm{and} \qquad
\theta^d_t \equiv d^{1/2} \, \bra{ \XX^d_{d \times t}, \UU^d_{d \times t}} / \| \XX^d_{d \times t} \|.
\end{align*}
Note that time has been accelerated by a factor $d$. The process $\theta^d_t$ describes the dot product between the position $X^d$ and the direction $U^d$,  scaled by a factor $d^{1/2}$ in order to observe a non-degenerate limiting process. It{\^o}'s lemma, neglecting terms of order $1/d$, directly shows (after straightforward algebra) that the Markov process $(R^d_t, \theta^d_t)$ has a generator $\cL^{\epsilon}$ that reads
\begin{align}\label{eq.fast.slow}
\cL^{\epsilon} = 
\epsilon^{-1} \, \cL^{(H)} + \epsilon^{-1} \, \frac{R}{\sigma^2} \, \theta_+ \, \cF + 
\epsilon^{-2}  \underbrace{ \left\{ \frac{1}{\sigma} \, \cL^{(J)} + \frac{\kappa}{2} \, \cL^{(K)}\right\}}_{ \cL^{(\fast)} } 
\end{align}
with the standard multiscale expansion notation $\epsilon = 1/\sqrt{d}$, generators $\cL^{(J)}$ and $\cL^{(K)}$ defined in Equation \eqref{eq.generator.fast} and
\begin{align*}
\cL^{(H)} \phi =\Big( \theta \, \partial_R - \frac{R}{\sigma^2} \, \partial_\theta \Big) \, \phi
\end{align*}
It is important to note that the generator $\cL^{(\fast)}$ describes the dynamics of a Markov process that is ergodic with respect to the standard centred Gaussian distribution $\mathrm{G}(\md \theta)$.
The dynamics described by the generator \eqref{eq.fast.slow} is a {\it fast-slow} systems with slow variable $R$ and fast variable $\theta$. The effective dynamics of the slow variable $R$ as $\epsilon \to 0$, or equivalently as $d \to \infty$, can be obtained with a standard multiscale expansion  \citep{papanicolaou1977martingale,weinan2011principles}, as described for example in Chapter $11$ of \cite{pavliotis2008multiscale}. One seeks a solution $\phi^{\epsilon}(t,R,\theta)$ to the backward Kolmogorov equation $(\partial_t - \cL^{\epsilon}) \phi^{\epsilon} = 0$ expressed as $\phi^\epsilon(t,R,\theta) = \phi(t,R) + \epsilon \, A(t, R, \theta) + \epsilon^2 \, B(t,R,\theta) + \mathcal{O}(\epsilon^3)$ in order to obtain the generator $\cL$ describing the leading term $\phi$, i.e. $(\partial_t - \cL) \phi = 0$. Expanding the Kolmogorov Equation $(\partial_t - \cL^{\epsilon}) \phi^{\epsilon} = 0$ in powers of $1/\epsilon$ shows that
\begin{align}\label{eq.multiscale}
\begin{aligned}
\mathcal{O}(\epsilon^{-2}): &\qquad  \cL^{(\fast)} \phi = 0 \\
\mathcal{O}(\epsilon^{-1}): &\qquad  \cL^{(\fast)} A = -\cL^{(H)} \phi \\
\mathcal{O}(1): &\qquad \partial_t \phi = \big( \cL^{(H)} + \sigma^{-2} \, R \, \theta_+ \cF \big) A + \cL^{(\fast)} B 
\end{aligned}
\end{align}
Equation $\cL^{(\fast)} \phi = 0$ is immediate since $\phi$ does not depend on the variable $\theta$.
Furthermore, we have that $\cL^{(H)} \phi=\theta \partial_R \phi(t,R)$. Consequently, it follows from the second equation of \eqref{eq.multiscale} that $A = c(t,R) + g(\theta) \, \partial_R \phi(t,R)$ where $g: \bR \to \bR$ is solution of the Poisson equation $\cL^{(\fast)} g = -\theta$ and $c(t,R)$ is a function that does not depend upon $\theta$. For an arbitrary function $(t, R,\theta) \mapsto h(t, R,\theta)$, we now use the standard notation $\bra{h}_{\mathrm{G}}$ to denote the operation of averaging out the fast variable $\theta$ over the standard centred Gaussian distribution $\mathrm{G}(\md \theta) = (2 \pi)^{-1/2} e^{-\theta^2/2} \, \md \theta$, i.e. $\bra{h}_{\mathrm{G}}(t,R) \equiv \int_{\bR} h(t, R,\theta) \, \mathrm{G}(\md \theta)$. We have that $\bra{\cL^{(\fast)} B}_{\mathrm{G}} \equiv 0$ so that the second equation in \eqref{eq.multiscale} leads to
\begin{align*}
\begin{aligned}
\partial_t &\phi(t,R)
= \bra{\cL^{(H)}A + \frac{R}{\sigma^2} \theta_+ \cF A}_{\mathrm{G}} \\
&= \bra{\theta \, g(\theta)}_{\mathrm{G}} \, \partial_{RR} \phi(t,R) 
+ \sigma^{-2} \, R  \, \bra{-g'(\theta) + \theta_+ \, \cF g(\theta)}_{\mathrm{G}}\, \partial_R \phi(t,R) \equiv \cL \phi(t,R).
\end{aligned}
\end{align*}
Algebra and an integration by part (i.e. Stein's Lemma) show that $\bra{-g'(\theta) + \theta_+ \, \cF g(\theta)}_{\mathrm{G}}$ also equals 
$2 \, \bra{\theta \, g(\theta)}_{\mathrm{G}}$. Consequently, we have that
\begin{align}\label{eq.OU.limit.kolmogorov}
\partial_t \phi(t,R)
&= \bra{\theta \, g(\theta)}_{\mathrm{G}} \, \BK{-\frac{R}{\sigma^2/2} \, \partial_R + \partial_{RR} } \phi(t,R) 
\end{align}
Since $g$ is solution of the Poisson equation $\cL^{(\fast)}g(\theta) = -\theta$, the quantity $\bra{\theta \, g(\theta)}_{\mathrm{G}}$ also describes the following asymptotic variance,
\begin{align*} 
V_\sigma(\kappa) = \lim_{T \to \infty} \; \Var
\BK{ \frac{1}{\sqrt{2 \, T}} \, \int_0^T \, \theta^{\kappa}_t \, \md t },
\end{align*}
where $\theta^{\kappa}_t$ is the Markov process with generator $\cL^{(\fast)} \equiv \frac{1}{\sigma} \, \cL^{(J)} + \frac{\kappa}{2} \, \cL^{(K)}$. Indeed, Equation \eqref{eq.OU.limit.kolmogorov} is the Kolmogorov backward equation associated to the Ornstein-Uhlenbeck
\begin{align*}
d \RR_t = - 2 \, \sigma^{-2} \, V_\sigma(\kappa) \, \RR_t \, dt + \sqrt{2 \, V_\sigma(\kappa)} \, dW,
\end{align*}
which concludes the proof of Proposition \ref{prop.diffusion.limit}.

\section{High-dimensional behaviour for finite $\delta$ and non-isotropic target}
\subsection{Proof of Theorem \ref{thm.noniso}}
In this section, we use the notation $\Rightarrow$ to denotes convergence in distribution.
Since $U$ is a uniformly random unit vector we may write it as $U=Z/\|Z\|$, where $Z$ is a vector of independent standard Gaussians. Further, since $\|Z\|/\sqrt{d}\rightarrow 1$ in probability as $d\rightarrow \infty$, we henceforth substitute $U=Z/\sqrt{d}$. We also set $\xi_i=\gamma_i X_i$, so the $\xi_i$ are independent and identically distributed with a density of $\exp\{f(\xi_i)\}$. We use the shorthand of $g(x)=f'(x)$ and $h(x)=-f''(x)$, and we let $L$ be  the Lipschitz constant for $h$. Finally we set $\ell(X)\equiv\log \pi(X)$ and
define $B(X,U)\equiv\ell(X+\delta U)-\ell(X)$. Firstly,
\begin{align*}
  B(X,U)
  &=
  \delta U^\top \nabla \ell(X) +\frac{\delta^2}{2} U^\top \partial^2 \ell(X) U
  +\frac{\delta^2}{2} U^\top\left\{\partial^2\ell (X+\delta U)-\partial^2 \ell (X)\right\}U\\
  &=
  \frac{\delta}{\sqrt{d}}\sum_{i=1}^d \gamma_i Z_ig(\xi_i)
  -\frac{\delta^2}{2d}\sum_{i=1}^d\gamma_i^2 Z_i^2h(\xi_i)
  +\frac{1}{2\sqrt{d}}T_1^{(d)},
\end{align*}
where $|T_1^{(d)}|\le \delta^3/d \times L\sum_{i=1}^d\gamma_i^3|Z_i|^3\rightarrow \delta^3L \times \EE[\gamma^3]\EE[|Z_i|^3]<\infty$. Hence as $d\rightarrow \infty$, the Central Limit Theorem gives
\begin{align}\label{eqn.stdDiff}
  B(X,U)&\Rightarrow \mathsf{N}\left\{-\frac{1}{2}\delta^2 \betaE J,~\delta ^2\betaE J\right\}.
\end{align}
The result for $\alpha_{\mathrm{pu}}$ then follows from Proposition 2.4 of \cite{roberts1997weak}. 
The $i$th component of the gradient vector with respect to which a bounce might occur is
\begin{align}\label{eqn.ithgrad}
\begin{aligned}
  \partial_{x_i} \ell(X+\delta U)
  &=
  \gamma_i g(\gamma_i X_i + \gamma_i \delta U_i)
  =\gamma_i g(\xi_i+\frac{\delta}{\sqrt{d}}\gamma_i Z_i)\\
  &=
  \gamma_i g(\xi_i) -\frac{\delta}{\sqrt{d}}\gamma_i^2Z_ih(\xi_i)+R_i^{(d)},
\end{aligned}
\end{align}
where $|R_i^{(d)}| \le L\delta^2\gamma_i^3Z_i^2/d$.
Thus
$\|\nabla \ell(X+\delta U)\|^2  =   \sum_{i=1}^d \gamma_i^2 g(\xi_i)^2+\cO(1)$, so
\begin{align*}
  \|\nabla \ell(X+\delta U)\|^2/d
  &\rightarrow \EE[\gamma^2 g(\xi)^2]
=J\betaE.
\end{align*}
Also, by \eqref{eqn.ithgrad} and the central limit theorem,
\begin{align} \label{eqn.UdotGrad}
\begin{aligned}
  U^\top \nabla \ell(X+\delta U)
  &=
  \frac{1}{\sqrt{d}}\sum_{i=1}^d\gamma_iZ_ig(\xi_i)
  -\frac{\delta}{d}\sum_{i=1}^d\gamma_i^2Z_i^2h(\xi_i)
  +\cO(1/\sqrt{d})\\
  &\Rightarrow
  \mathsf{N}\left\{-\delta \betaE J, \betaE J\right\}.
\end{aligned}
\end{align}
Now $U-V=2 [U^\top \nabla \ell(X+\delta U)] /\|\nabla \ell(X+\delta U)\|^2\times \nabla \ell(X+\delta U)$, so from \eqref{eqn.UdotGrad} we have that
\begin{align*}
\begin{aligned}
  d \times \|U-V\|^2
  &=
  \frac{4d}{\|\nabla \ell(X+\delta U)\|^2}
  \times [U^\top \nabla \ell(X+\delta U)]^2\\
&\Rightarrow \frac{4}{J\betaE}  [\mathsf{N}\{-\delta \betaE J, \betaE J\}]^2,
\end{aligned}
\end{align*}
so that $\|U-V\|=\cO(1/\sqrt{d})$. Consequently, the quantity $\ell(X+\delta U + \delta V)-\ell(X+\delta U)$ equals
\begin{align*}
\delta V^\top \nabla \ell(X+\delta U))-\frac{1}{2}\delta^2 V^\top \partial^2 \ell(X+\delta U) V
+\frac{1}{2\sqrt{d}}T_2^{(d)},
\end{align*}
where $|T_2^{(d)}|\le\delta^3L\sum_{i=1}^d\gamma_i^3V_i^2|Z_i|=\mathcal{O}(1)$ by the Lipschitz condition on $f''$, the boundedness of $\EE[\gamma^3]$ and because $\|U-V\|^2=\cO(1/d)$ and $U=Z/\sqrt{d}$. Further, the quantity $-B(X,U)$ also reads
\begin{align*}
\ell(X)-\ell(X+\delta U)
&=
-\delta U^\top \nabla \ell(X+\delta U))-\frac{1}{2}\delta^2 U^\top \partial^2 \ell(X+\delta U) U+\cO(1/\sqrt{d}).
\end{align*}
Subtracting the two expressions and noting that the bounce is constructed so that $(U+V)^\top \nabla \ell(X+\delta U)=0$ yields
\begin{align}\label{eqn.size.of.refl.diff}
\begin{aligned}
  \ell(X+\delta U + \delta V)-\ell(X)&=
  \delta (U+V)^\top \nabla \ell(X+\delta U))
  +\frac{1}{2}\delta^2 U^\top \partial^2 \ell(X+\delta U) U\\
&\qquad  -\frac{1}{2}\delta^2 V^\top \partial^2 \ell(X+\delta U) V+\cO(1/\sqrt{d})\\
&= \frac{1}{2}\delta^2(U-V)^\top \partial^2 \ell(X+\delta U) (2U-(U-V))+\cO(1/\sqrt{d}).
\end{aligned}
\end{align}
Since $\|U\|=1$ and $\|U-V\|=\cO(1/\sqrt{d})$, the term to control in \eqref{eqn.size.of.refl.diff} is
\begin{align*}
  \delta^2|(U-V)^\top \partial^2 \ell(X+\delta U) U|
  &=
  2\delta^2 \frac{U^\top \nabla(X+\delta U)}{\|\nabla\ell(X+\delta U)\|^2}
  |U^\top \partial^2 \ell(X+\delta U) \nabla\ell(X+\delta U)|\\
  &\sim
  2\delta^2 \frac{|\mathsf{N}\{-\delta \betaE J, \betaE J\}|}{d J \betaE}
  |W^{(d)}|,
\end{align*}
where 
\begin{align*}
  W^{(d)}&\equiv\frac{1}{\sqrt{d}}\sum_{i=1}^dZ_i\gamma_i^3h(\xi_i+\gamma_i Z_i/\sqrt{d})\gamma_ig(\xi_i+\gamma_iZ_i/\sqrt{d})\\
  &=\frac{1}{\sqrt{d}}\sum_{i=1}^d\{Z_i\gamma_i^3h(\xi_i)\gamma_ig(\xi_i)+\cO(1/\sqrt{d})\}.
\end{align*}
But $\EE\{\sum_{i=1}^dZ_i\gamma_i^3h(\xi_i)g(\xi_i)\}=0$ so $W^{(d)}=\mathcal{O}(1)$, and, hence, $\ell(X+\delta U + \delta V)-\ell(X)=\cO(1/\sqrt{d})$. It follows that
\begin{align*}
\ell(X+\delta U)-\ell(X+\delta U + \delta V)
&=
\ell(X+\delta U)-\ell(X) -\{\ell(X+\delta U + \delta V)-\ell(X)\}
\rightarrow
B(X,U)
\end{align*}
in probability.
If there is a delayed-rejection event then the standard move must have been rejected and, for example, $B$ must be negative.
Let $\mathsf{DR}$ be the event that the standard move has been rejected and so a delayed-rejection step is being attempted. 
Let $f_{B}(b)$ be the \emph{a priori} density for $B$ at stationarity, and let $f_{B|\mathsf{DR}}(b)$ be the density conditional on there being a delayed-rejection event. Then
\begin{align} \label{eqn.condDens}
f_{B|\mathsf{DR}}(b)
=\frac{f_{B}(b)\left\{1-1\wedge \exp(b)\right\}}
  {\int_{-\infty}^\infty f_{B}(b)\left\{1-1\wedge\exp(b)\right\}\md b},
\end{align}
which is well-behaved and has no mass where $\alpha_{\text{dr}}(X^{(d)},U^{(d)})$ is undefined. In the limit as $d\rightarrow \infty$, $f_{B}(b)$ is the density of the Gaussian distribution in
\eqref{eqn.stdDiff}, and \eqref{eqn.condDens} gives the limiting conditional density. It follows from the Bounded Convergence Theorem that
\begin{align*}
  \alpha_{\text{dr}}(X,U)
  &=
  \EE\left[
    1\wedge
    \frac{1-1\wedge \exp\{\ell(X+\delta U)-\ell(X+\delta U + \delta V)\}}
         {1-1\wedge \exp\{\ell(X+\delta U)-\ell(X)\}}
         \exp\{\ell(X+\delta U + \delta V)-\ell(X)\}
         \mid \mathsf{DR}\right]\\
  &\rightarrow
  \EE\left[1\wedge 
    \frac{1-\{1\wedge \exp(B)\}}
         {1-\{1\wedge \exp(B)\}}
         \mid \mathsf{DR}
         \right] =1,
\end{align*}
as required.

\subsection{Simulation study varying $d$ for fixed $\delta$}
In dimension $d$ we explore a $\mathsf{N}\left\{0,\mathsf{diag}(\sigma_1^2,\dots,\sigma_d^2)\right\}$ target with $\gamma_i^2=1/\sigma_i^2=2i/(d+1)$. In this case, $J=1$ and, essentially, $\betaE=1$, so that for fixed $\delta$, Theorem \ref{thm.noniso} tells us that, asymptotically, we expect $\alpha_{\mathrm{pu}}\rightarrow 2\Phi(-\delta/2)$ and $\alpha_{\text{dr}}\rightarrow 1$. For each combination of $d\in\{5,10,20,50,100,200\}$ and $\delta \in\{0.1,0.2,0.5,1.0,2.0,3.0\}$ (except $d=5$ and $\delta=0.1)$ where the Monte Carlo relative error was large) three replicate runs were performed with $\kappa=1.0$. Empirical acceptance rates for the standard moves and for the delayed-rejection move were recorded for each run.

\begin{figure}[h]
  \begin{center}
\includegraphics[scale=0.4]{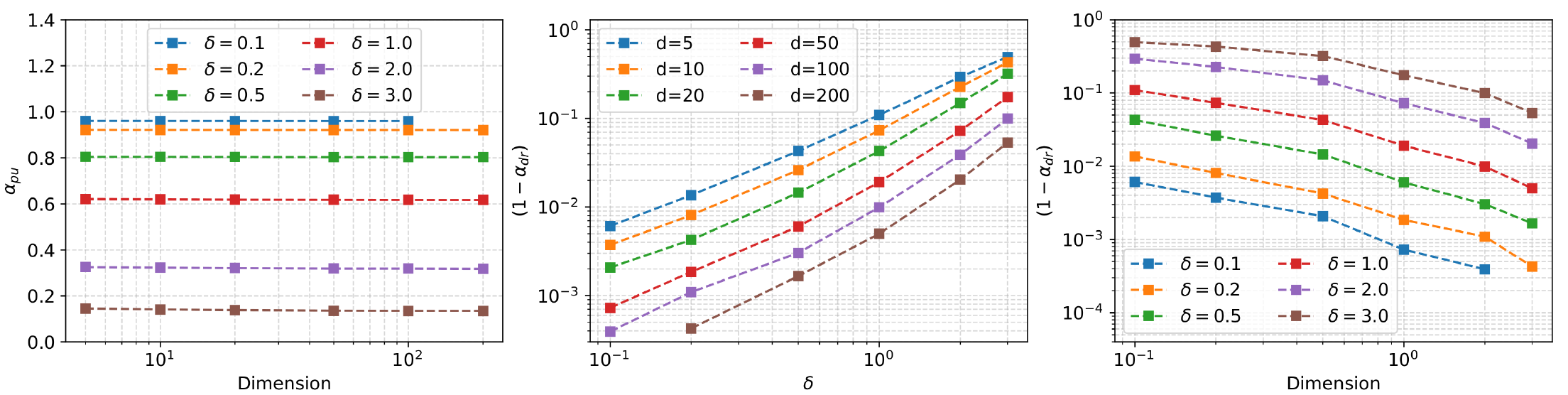}
    \end{center}
\caption{Left: $\alpha_{\mathrm{pu}}$ against $\log_{10}d$ split by $\delta$; centre: $\log_{10}(1-\alpha_{\text{dr}})$ against $\log_{10} d$ split by $\delta$; right: $\log_{10}(1-\alpha_{\text{dr}})$ against $\log_{10} \delta$ split by $d$.}
\label{fig.simstud.dinf}
\end{figure}
The left panel of Figure \ref{fig.simstud.dinf} plots $\alpha_{\mathrm{pu}}$ against the dimension $d$ for each fixed $\delta$ and demonstrates that for each fixed $\delta$, the acceptance rate $\alpha_{\mathrm{pu}}$ is insensitive to $d$, suggesting that the asymptotics are highly relevant even for very low dimension. The central and right panels plot $(1-\alpha_{\text{dr}})$ against the dimension $d$ and $(1-\alpha_{\text{dr}})$ against $\delta$ respectively, and suggests that, not only does $\alpha_{\text{dr}}\rightarrow 1$ as $d\rightarrow \infty$  but that, at least in this example, asymptotically, $1-\alpha_{\text{dr}}\propto \delta^2/d$.

\section{Further simulation studies}

\subsection{Convergence from the tails}
\label{sec.convfromtail}
We illustrate the contrast between the discrete bouncy particle sampler and Hamiltonian Monte Carlo on a fifty-dimensional target with a density described in Equation \eqref{eqn.powaniso}.
We first tuned each algorithm by starting at random positions of the form $(\sigma_1 z_1,\dots, \sigma_d z_d)\times r_{\star}$,
where $z\in \mathbb{S}^{d-1}$ is a uniformly random unit vector in $\mathbb{R}^d$. For Hamiltonian Monte Carlo we tried different integration times, $T$, and for each $T$, we followed the advice of \cite{beskos2013optimal}, noting that convergence to the optimal acceptance rate as dimension increases is slow, and tuned the number of leapfrog steps so as to achieve an acceptance probability of a little over 70\%. For the discrete bouncy particle sampler we tried different values for $\delta$, and for each $\delta$ we chose $\kappa$ so that, as suggested in Section 4.1, the mean dot product diagnostic is around 0.3-0.4. This suggested that $(T=2.0,L=4)$ for Hamiltonian Monte Carlo and $(\delta=2.0,\kappa=0.7)$ for the discrete bouncy particle sampler were reasonable tunings. Since, for a non-toy target in $d=50$, gradient evaluations will be much more costly than likelihood evaluations it is reasonable to a first approximation to assess efficiency by comparing the ratio of effective sample size to the average number of gradient evaluations per iteration. With $10^5$ iterations this is $\approx 1473$ for Hamiltonian Monte Carlo and $\approx 1014$ for the discrete bouncy particle sampler. However, the apparent relative success of Hamiltonian Monte Carlo disguises a serious underlying issue. 

We reran Hamiltonian Monte Carlo for $10^6$ iterations $40$ additional times with $X_0=\gamma(\sigma_1 z_1,\dots, \sigma_d z_d)\times r_{\star}$ for each $\gamma\in\{1.5,2.0.2.5,3.0\}$, with a new, independent $z$ vector on each of the $160$ occasions. On each occasion we counted the fraction of times where, by iteration $10^6$ the algorithm had ever had a value with $\|x\|_{\text{M}}\le r_{\star}$; i.e., the algorithm had reached the main posterior mass. The number of runs which converged by this measure were: $40/40~(\gamma=1.5)$, $36/40~(\gamma=2.0)$, $4/40~(\gamma=2.5)$ and $0/40~(\gamma=3.0)$; indeed, for every run with $\gamma=3.0$ the empirical acceptance rate was exactly zero. This fits with the known lack of geometric ergodicity of Hamiltonian Monte Carlo on light-tailed targets. By contrast, for the discrete bouncy particle sampler with $\gamma=3.0$, all $40$ runs converged within $1000$ iterations, and, indeed, $26$ of the runs converged within $300$ iterations.

In summary, on this occasion, when both algorithms were started from the main posterior mass, the discrete bouncy particle sampler was competetive with Hamiltonian Monte Carlo, though less efficient. However, because our algorithm depends on $\nabla \log \pi$ only through the unit vector, it is robust to large $\|\nabla \log \pi\|$, unlike Hamiltonian Monte Carlo.

\subsection{The Markov modulated Poisson process}

\begin{figure}[h]
  \begin{center}
    \includegraphics[scale=0.7]{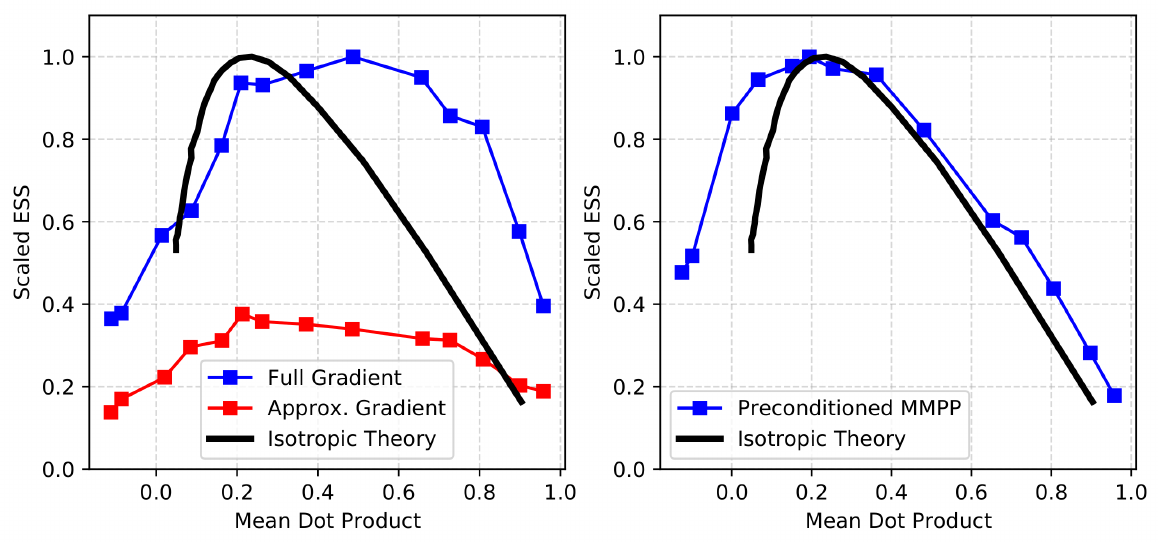}
    \end{center}
\caption{Scaled effective sample size as a function of the mean dot product for several values of $\kappa$, obtained from \DBPS runs of length $N=2\times 10^5$ applied to the logarithms of the MMPP parameters with no preconditioning and with $n_{\text{cpt}}=d=8$ and $n_{\text{cpt}}=3$ (Left), and with crude preconditioning and $n_{\text{cpt}}=d$ (Right).}
\label{fig.toy.mmpp}
\end{figure}

Finally, we consider a $k$-state, continuous-time Markov chain $Z_t$ started from
state $1$, and a Poisson process $N_t$ whose
rate $\lambda_t$ is a fixed function of $Z_t$. The doubly-stochastic process is parameterized by the rate matrix for the Markov chain, $Q$, and a vector of rates for the Poisson process, $\lambda$, where $\lambda_i,~(i=1,\dots,k)$ is the rate of $N_t$ when $Z_t=i$.

The event times of $N_t$ are observed over a time window
$[0,t_{\mathrm{end}}]$, but the behaviour of $Z_t$ is unknown, and we
wish to perform inference on $(Q,\lambda)$. Setting
$\Lambda=\mbox{diag}(\lambda)$, the likelihood for the number of
events $n$ and the event times $t_1,\dots,t_n$ is  \citep{FearnheadSherlock2006}:
\begin{align*}
  L(Q,\lambda;t)
  =
  \mathrm{e}' \exp[(Q-\Lambda)t_1]\Lambda \exp[(Q-\Lambda)(t_2-t_1)]\Lambda \dots
  \Lambda\exp[(Q-\Lambda)(t_{\mathrm{end}}-t_n)] 1,
\end{align*}
where $1$ is the $k$-vector of ones and $\mathrm{e}'=(1,0,\dots,0)$.
  We simulated a dataset using a cyclic four-state Markov chain for a
  $200$-second time window with $Q$ parameters of:
  $Q_{12}=Q_{23}=Q_{41}=1.0,~Q_{34}=0.25$ and all other
  off-diagonal rates set to zero. The rate parameters were
  $\lambda_1=20.0, \lambda_2=5.0,~\lambda_3=1.0$ and $\lambda_4=10.0$. We
  then conducted inference on the natural logarithm of each parameter
  that was not systematically zero,
  placing independent $N(0,2^2)$ priors on each of these. 

  Code was written in \texttt{C++} where auto-differentiation was not available for general matrix
  exponentials, and so numerical differentiation via centred differences was used (the cheaper, first-order Euler approximation led to precision problems). We applied
  the \DBPS for $2\times 10^5$ iterations for a number of $\kappa$ values and repeated this but evaluating only
  $n_{\text{cpt}}=3$ 
  randomly-orientated components of the eight-dimensional gradient vector on each
  delayed-rejection step. Figure \ref{fig.toy.mmpp} plots scaled effective sample size against $\kappa$ and suggests that the optimal mean dot product is around $0.5$ when all gradient components are used and around $0.2$ when three random components are used. The optimal effective sample size in the latter case is around $3/8$ of the former; since the number of gradient calculations during a bounce is also reduced by $3/8$ this suggests no loss in overall efficiency. When all components are used, tuning to a dot product of $0.2$ brings only a $10\%$ reduction in effective sample size. The posterior variance matrix has a condition number of $49.2$, so, following typical practice, for each $\kappa$ the \DBPS was rerun using a crude preconditioning matrix (Section \ref{sec.precond}) of $M=\mbox{diag}(1/2,2,1,1,2,2,2,2)$. Although the effective variance matrix still has a condition number of $\approx 4.3$ the right panel of Figure \ref{fig.toy.mmpp} shows that the optimal choice of $\kappa$ is now $\approx 0.2$.

\end{document}